\documentclass[a4paper,fleqn,usenatbib]{mnras}
\usepackage[T1]{fontenc}
\usepackage{ae,aecompl}

\usepackage{graphicx}	
\usepackage{amsmath}	
\usepackage{amssymb}	
\usepackage{xspace}

\newcommand{\logten}{\ensuremath{\log_{10}}}
\newcommand{\reff}{R_\textrm{eff}}
\newcommand{\reffc}{R_\textrm{eff,c}}

\newcommand{\SExtractor}{\textsc{SExtractor}\xspace}
\newcommand{\PSFEx}{\textsc{PSFEx}\xspace}

\newcommand{\MegaPrime}{{MegaPrime}\xspace}

\newcommand{\SkyMaker}{\textsc{SkyMaker}\xspace}

\newcommand{\mauto}{\texttt{MAG\underline{~}AUTO}\xspace}

\newcommand{\spmod}{\texttt{SPREAD\underline{~}MODEL}\xspace}
\newcommand{\angstrom}{\mbox{\normalfont\AA}}

\title[The abundance of compact quiescent galaxies]{The abundance of compact quiescent galaxies since $z \sim 0.6$}
\author[Ald\'ee Charbonnier et al.]{Ald\'ee Charbonnier$^{1}$\thanks{E-mail: charbonnier@astro.ufrj.br}, Marc Huertas-Company$^{2,3}$, Thiago S. Gon\c{c}alves$^{1}$, Kar\'in 
\newauthor Men\'endez-Delmestre$^{1}$, Kevin Bundy$^{4}$, Emmanuel Galliano$^{1}$, Bruno Moraes$^{5}$,  
\newauthor Mart\'in Makler$^{6}$, Maria E. S. Pereira$^{6}$, Thomas Erben$^{7}$, Hendrik Hildebrandt$^{7}$, 
\newauthor Huan-Yuan Shan$^{7}$, Gabriel~B.~Caminha$^{8}$, Marco Grossi$^{1}$, Laurie Riguccini$^{1}$ \\
$^{1}$ Observat\'orio do Valongo, Universidade Federal do Rio de Janeiro, Ladeira Pedro Ant\^onio 43, Sa\'ude, Rio de Janeiro, RJ, \\
CEP 20080-090, Brazil\\ 
$^{2}$ LERMA, Observatoire de Paris, CNRS, Universit\'e Paris Diderot, Paris Sciences et Lettres (PSL) Research University, Universit\'e\\
Paris Sorbonne Cit\'e, 61 Avenue de l'Observatoire, F-75014 Paris, France\\
$^{3}$ Department of Physics and Astronomy, University of Pennsylvania,
Philadelphia, PA 19104, USA\\
$^{4}$ UC Santa Cruz, 1156 High Street, Santa Cruz, CA 95064, USA\\
$^{5}$ Dept. of Physics and Astronomy, University College London, London, WC1E 6BT, UK\\
$^{6}$ Centro Brasileiro de Pesquisas F\'isicas, Rua Dr. Xavier Sigaud 150, CEP 22290-180, Rio de Janeiro, RJ, Brazil\\
$^{7}$ Argelander-Institut f\"ur Astronomie, Auf dem H\"ugel 71, D-53121 Bonn, Germany\\
$^{8}$ Dipartimento di Fisica e Scienze della Terra, Universit\`a degli Studi
di Ferrara, Via Saragat 1, I-44122 Ferrara, Italy}
\date{Accepted XXX. Received YYY; in original form ZZZ}

\pagerange{\pageref{firstpage}--\pageref{lastpage}}

\pubyear{2016}

\begin{document}
\label{firstpage}
\maketitle

\begin{abstract}

We set out to quantify the number density of quiescent massive compact galaxies at intermediate redshifts. We determine structural parameters based on $i$-band imaging using the CFHT equatorial SDSS Stripe~82 (CS82) survey ($\sim 170$ sq. degrees) taking advantage of an exquisite median seeing of $\sim 0\farcs 6$. We select compact massive ($M_{\star}>5\times 10^{10}\,M_{\sun}$) galaxies within the redshift range of $0.2<z<0.6$. The large volume sampled allows to decrease the effect of cosmic variance that has hampered the calculation of the number density for this enigmatic population in many previous studies. We undertake an exhaustive analysis in an effort to untangle the various findings inherent to the diverse definition of compactness present in the literature. We find that the absolute number of compact galaxies is very dependent on the adopted definition and can change up to a factor of $>10$. We systematically measure a factor of $\sim 5$ more compacts at the same redshift than what was previously reported on smaller fields with HST imaging, which are more affected by cosmic variance. This means that the decrease in number density from $z \sim 1.5$ to $z \sim 0.2$ might be only of a factor of $\sim 2-5$, significantly smaller than what previously reported. This supports progenitor bias as the main contributor to the size evolution. This milder decrease is roughly compatible with the predictions from recent numerical simulations. Only the most extreme compact galaxies, with $\reff < 1.5 \times \left( M_\star/10^{11}\,M_{\sun} \right)^{0.75}$ and $M_\star > 10^{10.7}\,M_{\sun}$, appear to drop in number by a factor of $\sim 20$ and hence likely experience a noticeable size evolution. 

\end{abstract}




\section{Introduction}

How galaxies form and evolve through cosmic time is one of the key questions of modern astronomy. In the context of hierarchical formation of structures in a cold dark-matter universe, less massive dark matter haloes form first, and then accrete to form more massive ones \citep[e.g.,][]{Diemand2011}. The star formation activity, however, is not simply proportional to the halo mass \citep[e.g.,][]{Somerville2015} nor constant over cosmic times \citep{Madau1998,Madau2014}. Hydro-cosmological simulations predict that massive galaxies tend to quench their star formation earlier and faster than less massive ones \citep{Zolotov2015}. Based on the Sloan Digital Sky Survey (SDSS, \citealt{York2000}) \citet{Citro2016} show that early-type massive galaxies follow an anti-hierarchical evolution (downsizing, \citealt{Cowie1996}), i.e. massive galaxies form and quench earlier. The star formation histories of the most massive galaxies reveal that they should have been formed by a vigorous star formation event, and have a compact configuration by $z\sim 2-3$. An important population of passively evolving massive galaxies is found to be already in place at $z\sim 2$ when the universe was only $\sim3$~Gyrs old \citep{Cimatti2004, Daddi2005, Trujillo2006, Damjanov2009, Whitaker2012, Cassata2013, HuertasCompany2013, HuertasCompany2015}. These high-redshift massive quiescent galaxies have been shown to be 3 to 5 times more compact than their local counterparts and have come to be commonly referred to as `red-nuggets' \citep{Daddi2005, Trujillo2006, Longhetti2007, Cimatti2008, VanDokkum2008, VanDokkum2010, Damjanov2009, Damjanov2011, Newman2010, Bruce2012, Ryan2012, Cassata2013, VanderWel2014}.

It remains to be understood how these massive compact galaxies formed: what are their star-forming progenitors and what are the quenching processes involved to turn them into quiescent galaxies? Comparing the number density, stellar mass and size of the population of submillimeter galaxies (SMG) at high redshifts ($z \gtrsim 3$) to those of the population of compact massive quiescent galaxies at $z\sim 2$ in the COSMOS field, \citet{Toft2014} proposed a direct evolutionary connection between these two extreme galaxy types. Some of the SMGs in the process of quenching could be the ones observed by \citet{Barro2013} and \citet{VanDokkum2015} who have identified in the CANDELS fields high-redshift ($z>2$) star-forming progenitors with similar sizes ($\sim 1$\,kpc), masses ($\sim 10^{11}\,M_{\sun}$) and number densities to those of quiescent compact galaxies. Measuring star formation rates and gas content of their star-forming candidates, these authors underline the presence of a gas disc with sizes ranging from $\sim 0.2$ to 10\,kpc. Compaction is expected to be associated with a quenching episode \citep{Dekel2014, Zolotov2015}, where the more compact the star-forming massive galaxies become, the higher the probability is of them being quenched \citep{Williams2015, VanDokkum2015}. Although many of these works call for rapid quenching to form these compact massive quiescent galaxies, there is currently no consensus regarding the quenching mechanism. \citet{Barro2013}, \citet{ForsterSchreiber2014} and \citet{VanDokkum2015} find that close to half of the star-forming progenitors host active galactic nuclei (AGNs), but there is yet no direct evidence that the AGN is able to drive away the gas by itself \citep{Zolotov2015}. Environmental quenching \citep{Peng2010} may also be at play. Alternative models suggest that the origin of compact galaxies and subsequent size evolution might simply be a consequence of the size evolution of star-forming galaxies coupled with progenitor bias effects \citep[e.g.,][]{Carollo2013, Lilly2016}.

Recent cosmological hydrodynamic simulations have contributed to our understanding of the evolution and formation of these compact massive galaxies. Taking into account a number of physical processes (e.g., gravity, hydrodynamics, gas cooling, star formation, stellar evolution, supernova and black hole feedback)  and using subgrid models for feedback processes (e.g., \textsc{EAGLE}, \citealt{Schaye2015, Crain2015} and \textsc{ILLUSTRIS}, \citealt{Vogelsberger2014}) they seek to reproduce observed properties. Compact massive quiescent galaxies have indeed been identified in simulations at high redshifts \citep{Wellons2015, Zolotov2015, Furlong2015}, allowing us to trace back their formation history. \citet{Zolotov2015} find that the compaction of star-forming discs must be driven at a rate faster than the star formation rate. The compaction might be due to mergers (both minor and major) in concert with violent disc instabilities \citep[e.g.,][]{Dekel2014}. The onset of quenching happens when the galaxy is at its maximum compactness, due to gas depletion. Simulations show therefore that very compact passive galaxies are formed at high redshifts through dissipative process associated with gas inflow into the galaxy central region. As the efficiency of the quenching and compaction processes are correlated with the abundance of cold dense gas in the universe, the production of compact massive quiescent galaxies is expected to decrease with cosmic time. An alternative scenario to explain the shut-down of the star formation invokes the mechanism of `halo quenching' \citep[e.g.,][]{Zu2016}, in which the mass of the dark matter halo is the main driver for triggering the quenching. However, \citet{Woo2015} showed that galaxy compactness is playing a non-negligible role for satellite galaxy quenching.

Both observations \citep[e.g.,][]{VanderWel2014} and cosmological hydrodynamic simulations \citep[e.g.,][]{Wellons2016} agree on the fact that massive early-type galaxies are smaller in median size at higher redshifts. This is partly due to the so-called progenitor bias, as star-forming galaxies have larger sizes at later times \citep{Mo1998}. An intrinsic growth of individual galaxies is expected and observed in numerical simulations, due e.g., to: (i) radial migration of stars, (ii) addition of mass to the outer region by mergers or accretion, and (iii) renewed star formation at larger radii. These channels of size evolution have been suggested to potentially lead the original compact massive quiescent galaxies to become the bulges of local galaxies \citep{Graham2015,DelaRosa2016}. These processes are however stochastic: one expects to see a decrease in numerical density of compact massive quiescent galaxies with redshift, leaving some relic candidates untouched. \citet{Wellons2016} and \citet{Furlong2015} studied the evolution of massive compact quiescent galaxies since high redshift using the \textsc{ILLUSTRIS} and the \textsc{EAGLE} simulations, respectively. On the one hand, \citet{Furlong2015} show that among the original population of compact massive and passive galaxies at $z=2$, 15\% remain central cores, 25\% become satellites, and 60\% merge with more massive systems at low redshifts. On the other hand, \citet{Wellons2016} observe that 14\% were consumed and accreted by more massive galaxies, 6\% experienced a major merger event and are partially disrupted, 49\% remained as the core of a more massive descendant, and 31\% remained untouched. The growth in size of galaxies with cosmic times is accompanied by a growth in mass that is dominated by ex-situ stars. \citet{Wellons2016} underline that even the undisturbed compact massive sample grows in mass and size. They also confirm the result obtained by \citet{Oser2012}, showing that the dominant accretion mode for simulated massive galaxies from $z\sim 2$ to present time is minor mergers with a mass-weighted mass ratio of 1:5. Despite the mean growth in size and mass of the high redshift population of massive compact quiescent galaxies, candidates that did not yet accrete large numbers of external stars and remained compact are expected to be found in the local universe.

The current picture of (and search for) compact massive quiescent galaxies in the local universe is quite unclear. Depending on the definition of compactness and on the nature of the dataset, results are dramatically different. \citet{Trujillo2009} find almost no candidates using the SDSS DR6 NYU value-added galaxy catalogue (NYU VAGC, $z<0.2$), where they classify $0.03$\% of the population of massive ellipticals as relics. \citet{Valentinuzzi2010a} find, in turn, a large sample in the WIde-field Nearby Galaxy-cluster Survey ($0.04<z<0.07$) where $22$\% of the cluster members with masses in the interval $3\times 10^{10} < M_\star/M_{\sun} < 4\times 10^{11}$ are red nugget candidates. Based on the spectroscopic sample of SDSS DR7 \citet{Taylor2010} find a marked dearth of massive quiescent galaxies in the local universe ($0.066 < z < 0.12$) that are as compact as those at high redshifts. Adopting the same definition for compact relics as \citet{Valentinuzzi2010a}, \citet{Poggianti2013a} look for candidates in the field in the context of the Padova-Millennium Galaxy and Group Catalogue ($0.03 < z < 0.11$), and find three times less compact galaxies in the field than in cluster environments. \citet{Trujillo2014} identify a nearby galaxy, NGC 1277, as being one representative of the massive compact relics, with a mass of $1.2 \pm 0.4 \times 10^{11}\,M_{\sun}$ and a mean age of 12\,Gyr. Based on the same catalogue as \citet{Trujillo2009} but defining differently the compact relics, \citet{PeraltaDeArriba2016} find that galaxy clusters might be the preferred environments to find compact relics. 

Building a coherent picture that brings together the set of consistent observations at high redshifts with the discrepant ones locally calls for an analysis based on an intermediate redshift sample, as well as a common definition of the so-called local red nuggets. Such an analysis would provide insights on the evolution processes of the compact massive quiescent galaxy population over cosmic times. The main challenge of intermediate redshift studies is the chosen compromise between the area surveyed and the quality of the images, in order to have enough statistics on the one hand and to be able to disentangle stars from compact galaxies on the other. Space-based surveys sample limited volumes but benefit from exquisite image resolution, whereas ground based surveys can probe larger volumes but suffer from observational seeing limitations. Recent works using space-based data from HST found candidates for compact massive quiescent galaxies at intermediate redshifts in the COSMOS field \citep{Carollo2013,Damjanov2015} and in the ESO Distant Clusters Survey \citep{Valentinuzzi2010b}. Concerning ground based images, \citet{Damjanov2013} confirmed that non resolved galaxies of SDSS that are identified by their spectra present properties of compact quiescent candidates. \citet{Damjanov2014} extended this approach to data from the Baryon Oscillation Spectroscopic Survey (BOSS, \citealt{Eisenstein2011}) and derived the density evolution of compact massive candidates at intermediate redshifts, although with large error bars. Recently, \citet{Tortora2016} made use of the first and second releases of the ESO Public optical Kilo Degree Survey (KiDS) $-$ covering a region of $156$\,deg$^2$ in four bands $-$ and applying similar definitions as in \citet{Trujillo2009}, identified a population of compact relic candidates at intermediate redshifts; the number density of these objects appears to stay constant towards lower redshifts within the measured uncertainties. Although the authors do not observe any candidates at $z<0.2$, they attribute this to environment effects.

In the present analysis we look for compact massive quiescent galaxies at intermediate redshifts (from $z=0.2$ to $z=0.6$) in the so-called Stripe 82 region. Thanks to uniform, deep, multiwavelength and weak lensing quality data over this large equatorial stripe, we are currently into a privileged position to look for the population of compact candidates. In this work, we adopt a spatially flat cosmological model with $\Omega_M = 0.3$, $\Omega_\Lambda = 0.7$ and $H_0=70$\,km\,s$^{-1}$\,Mpc$^{-1}$. Magnitudes are quoted in the AB system.

\section{Identifying compact candidates in Stripe 82}

The so-called Stripe~82 is an equatorial stripe of $\sim 250$\,deg$^2$ in the southern Galactic cap. It has been observed by the SDSS repeatedly as part of a supernova survey \citep[e.g.,][]{Abazajian2009} significantly increasing the depth $-$ compared to single-pass SDSS data $-$ in the survey footprint for all {\it ugriz } optical bands, ($i\sim 24.3$ for point like sources at $3\sigma$, \citealt{Annis2014, Jiang2014, Fliri2016}). Following these observations, this field has benefited from a wide coverage from radio wavelengths to X-rays and has thus become a favourite field for large-scale multiwavelength studies. We list here part of Stripe~82 observations to give an idea of the continuous inflow of new data: deep radio data by the Karl G. Jansky Very Large Array (VLA \citealt{Hodge2011,Mooley2016,Heywood2016}); microwave from the Atacama Cosmology Telescope (ACT, \citealt{Swetz2011}); submillimeter from the Herschel satellite \citep{Viero2014}; infrared (IR) from the Spitzer-IRAC instrument \citep{Papovich2016,Timlin2016}; near-infrared (NIR) from the Wide-Field Infrared Survey Explorer (WISE, \citealt{Wright2010}), the UKIRT Infrared Deep Sky Survey (UKIDSS, \citealt{Lawrence2007}) and from a joint VISTA-CFHT survey (Geach et al. in preparation); optical imaging from SDSS, the Subaru Hyper Suprime Cam \citep{Miyazaki2012, Aihara2017}, $i$-band CFHT data from CS82 (Kneib et al. in preparation), 12 optical bands from the S-PLUS survey (C. Mendes de Oliveira et al. in preparation), and X-rays from Chandra and XMM-Newton data \citep{LaMassa2016, Rosen2016}. In terms of spectroscopy, Stripe~82 has been targeted by various surveys, including the SDSS-III Baryon Oscillation Spectroscopic Survey \citep[BOSS,][]{Dawson2013} and SDSS-IV \citep{SDSSDR13}, the WiggleZ Dark Energy Survey \citep{Drinkwater2010}, the 2dF Galaxy and QSO Redshift Surveys \citep{2dFGRS,2QZ}, the 6dF Galaxy Survey \citep{6dF}, the Deep extragalactic Evolutionary Probe \citep[DEEP2,][]{DEEP2}, the VIMOS VLT Deep Survey \citep[VVDS,][]{Garilli2008}, the VIMOS Public Extragalactic Redshift Survey \citep[VIPERS,][]{VIPERS}, and the PRIsm MUlti-object Survey \citep[PRIMUS,][]{Coil2011}. Being observable from both south and north hemispheres, this area is becoming a preferred field for calibration purposes of large photometric surveys such as the Dark Energy Survey \citep{DES2016} and the Large Synoptic Survey Telescope project \citep{LSST2009}.

\subsection{Datasets and catalogues\label{sec:datasets}}

Different photometric redshift estimators have been applied to SDSS-Stripe~82 catalogues. \citet{Bundy2015} find that the best performances are obtained both using the red-sequence Matched-filter Probabilistic Percolation algorithm (redMaPPer, \citealt{Rykoff2014}), and a neural network approach as done by \citet{Reis2012} with ANNz \citep{Collister2004}. When photo-z's are not available from these catalogues, EAZY (for Easy and Accurate Redshifts from Yale, \citealt{Brammer2008}) estimates are used. The details of the photo-z catalogue are explained in \citet{Bundy2015}. In this work we give preference to spectroscopic over photometric redshift when available.

In the NIR, UKIDSS \citep{Lawrence2007} targeted the Stripe~82 region, reaching $Y\sim 20$. We use the stellar masses and K-corrected colours from \citet{Bundy2015}\footnote{\url{http://massivegalaxies.com}}, which were obtained after applying the SYNthetic aperture MAGnitudes software ({\sc SYNMAGs}, \citealt{Bundy2012}) to match the photometry of SDSS coadd with UKIDSS, assuming a \citet{Chabrier2003} initial mass function. 
Considering that stellar mass estimates are more robust when using NIR data \citep{Courteau2014}, only objects that have at least one detection in one of the UKIDSS NIR bands ($Y$, $J$, $H$, and $K$) are considered for our analysis.

We derived the morphological parameters based on the CFHT/\MegaPrime Stripe~82 (CS82) survey. CS82 has been designed to provide high quality {\it i}-band imaging for a large fraction of the Stripe~82 region, suitable for weak lensing measurements (Kneib et al. in preparation). Due to the lensing specifications, an excellent image quality to a medium depth is required: the median seeing is $\sim 0\farcs 6$ and the limiting magnitude $mag_{\rm lim}\approx 24$ for a point-like source detection at 5$\sigma$. We run \SExtractor\footnote{\url{http://www.astromatic.net/software/sextractor}} \citep[v2.18.8,][]{Bertin1996} and \PSFEx\footnote{\url{http://www.astromatic.net/software/psfex}} \citep[v2.15.0,][]{Bertin2011} codes to characterize the morphology of all objects detected on the coadded images. Both codes have been designed to be run on large area images. \SExtractor provides the morphological parameters by fitting defined brightness profiles, taking into account the point spread function (PSF) estimated by \PSFEx. \PSFEx and \SExtractor were compared to the {\sc DAOPHOT} and {\sc ALLSTAR} software packages on simulated images by \citet{Annunziatella2013}. They find that \PSFEx performs accurate PSF modeling. Both codes were also used by \citet{Desai2012} to produce a PSF corrected model-fitting photometry catalogue of the Blanco Cosmology Survey.

The brightness profile of a galaxy is commonly fitted by the general S\'ersic parametrization which depends on the S\'ersic index $n_{\textrm{ser}}$. We fit four brightness profiles to the data: (1) a de Vaucouleurs ($n_{\textrm{ser}} = 4$) and (2) an exponential ($n_{\textrm{ser}} = 1$) profile, which respectively suit the brightness profiles of early and late-type galaxies, (3) a general S\'ersic profile and (4) a sum of a de Vaucouleurs and an exponential one. The morphological catalogue for the entire CS82 sample is based on the same approach (for details see Moraes et al. in preparation). PSF extraction is the cornerstone of our search of compact elliptical galaxies. \PSFEx performances are assessed in Moraes et al. (in preparation) by comparing the galaxy ellipticities recovered by \PSFEx and \SExtractor with the ellipticities obtained with the {\it lensfit} Bayesian shape measurement algorithm \citep{Heymans2012, Miller2013}, which we consider as a benchmark. For the ellipticities of PSF-sized galaxies we find a mean bias of $\sim 0.01 \pm 0.05$.  For galaxies with sizes below the PSF, \PSFEx and \SExtractor still allow to recover the correct ellipticities, but within a larger error bar: $\sim 0.04\pm 0.12$. Size estimates based on \PSFEx and \SExtractor for galaxies whose angular size is close to the PSF are discussed in section~\ref{sec:sizecheck}.

We measure the effective radii of the galaxies in the $i$-band. The pivot wavelength of the $i$ band corresponds to restframe wavelengths of 635~nm and 477~nm at redshifts $z=0.2$ and $z=0.6$, respectively. According to \citet{Kelvin2012}, the morphological K-correction resulting within this wavelength range is of the order of $\sim 0.1$~kpc for a spheroidal galaxy with an effective radius of 1~kpc, and of the order of $\sim 0.3$~kpc for an effective radius of 3~kpc. This 10\% effect on the effective radius is of the order of the measurement error from \SExtractor.

\subsection{Sample selection \label{sec:samplesel}}

\begin{figure}
  \includegraphics[width=84mm]{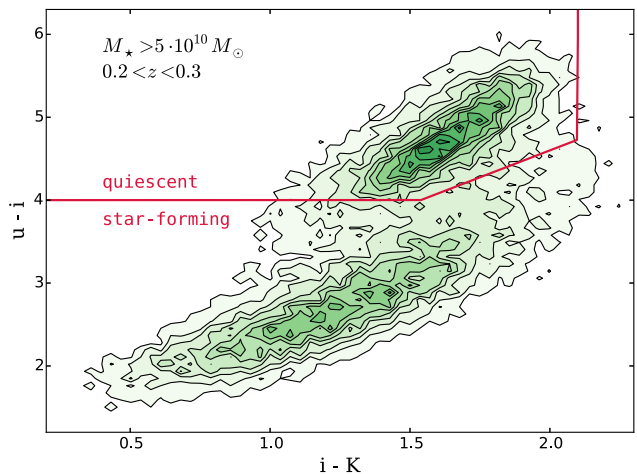}
  \caption{Colour-colour diagram illustrating the bimodal distribution of galaxies: star-forming late-types and quiescent early-types occupy two distinct regions within this diagram. The red line shows the adopted separation between the two populations in order to select quiescent objects. The green regions represent $\sim60,000$~objects in the redshift range $0.2 < z < 0.3$, where no stellar mass cut has been applied. The colours have been K-corrected.}
  \label{fig:quiescent}
\end{figure}

The present analysis focuses on the population of massive passive galaxies at intermediate redshifts. The way in which compact massive quiescent galaxies are selected, in particular the definition of compactness and of the lower limit used for the mass, has a great influence on the derived sample \citep[e.g.,][]{Damjanov2015}. We adopt different definitions following a set of different selection criteria that have been used by other authors in an effort to make reliable  comparisons with a broad range of previous studies \citep{Quilis2013, Carollo2013, VanderWel2014, VanDokkum2015}. We describe below the adopted cuts in redshift ($z$), stellar mass ($M_{\star}$) and colours:

\begin{enumerate}

\item $0.2 < z < 0.6$. We are interested in the evolution of the population of massive galaxies between high and low redshifts. The limits have been set to fill the gap  between these two regimes, following \citet{Damjanov2013}. Moreover our photometric redshift catalogue is reliable out to $z \sim 0.6$.

\item $\logten(M_{\star}/M_{\sun}) > \logten(M_\textrm{min}/M_{\sun})$. We note that to avoid contamination from star-forming galaxies, \citet{Moresco2013} recommend a cut of $\logten(M_\textrm{min}/M_{\sun}) = 10.75$, independently of the selection criteria to separate passive galaxies. For this reason we have decided not to include a comparison with \citet{Barro2013} and \citet{Poggianti2013a}, as they selected massive galaxies with a minimal mass of $\logten(M_\textrm{min}/M_{\sun}) = 10$ and $10.3$, respectively. In the present analysis, we follow the definitions of \citet{Quilis2013}: $\logten(M_\textrm{min}/M_{\sun}) = 10.9$ (corresponding to $M_\textrm{min} = 8\times 10^{10}\,M_{\sun}$); \citet{Carollo2013}: $\logten(M_\textrm{min}/M_{\sun}) = 10.5$; \citet{VanderWel2014}: $\logten(M_\textrm{min}/M_{\sun}) = 10.7$ and \citet{VanDokkum2015}: $\logten(M_\textrm{min}/M_{\sun}) = 10.6$. 

\item The use of the colour bimodality to separate early-type quiescent galaxies from late-type star-forming ones has been first underlined by \citet{Strateva2001} and \citet{Baldry2004}. In more recent analyses, \citet{Wuyts2007}, \citet{Williams2009}, \citet{Whitaker2011} and \citet{Muzzin2013} have worked in the rest frame colours $u-V$ vs. $V-J$. We follow their approach, applying adapted cuts for each redshift bin, defining four slices of $\Delta z =0.1$ between $z=0.2$ and $z=0.6$. To obtain the rest frame colours, a K-correction has been applied following the methodology of \citet{Chilingarian2010}. As an example, we show in Figure~\ref{fig:quiescent} how quiescent galaxies within the redshift bin $0.2 < z < 0.3$ are selected based on a cut defined to roughly match the local minima  between the two peaks in the $u-i$ vs. $i-K$ colour-colour galaxy distribution. For the redshift range $0.3 < z < 0.6$, we choose $g-i$ for the y-axis. This ensures these colours encompass the rest-frame 4000\,$\angstrom$ break, which strongly correlates with the age of the stellar population \citep[e.g.,][]{Martin2007, Goncalves2012}. According to the availability of the NIR bands from UKIDSS, we changed the x-axis to $i-H$ or $i-Y$.
\end{enumerate}

\subsection{Star/galaxy separation}

Considering that we are searching for compact galaxies that may resemble point-like sources, it is of major importance to have an accurate star/galaxy discriminator. We base our star/galaxy discrimination on CS82 data for a first sorting based on a morphological approach using \SExtractor. We further refine our discrimination with colour information from SDSS and UKIDSS.

\begin{figure}
	\includegraphics[width=\columnwidth]{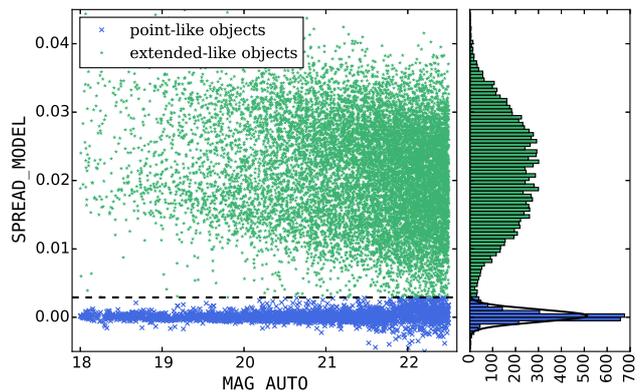}
	\caption{Distribution of the \spmod \SExtractor parameter as a function of the Kron magnitude \mauto for one tile of the CS82 survey. A gaussian function has been fitted to the \spmod histogram of the stellar branch, and values greater than three sigma above that are defined as extended (green stars).}
	\label{fig:spreadmodel}
\end{figure}

\SExtractor in its model-fitting feature provides the \spmod parameter that estimates the similarity of the brightness profile of an object to the image PSF (see eq.~5 of \citealt{Desai2012}). The distribution of \spmod as a function of the \SExtractor \mauto Kron magnitude is shown in Figure~\ref{fig:spreadmodel} for one coadded pointing (which we refer to as {\it tile}) of the CS82 survey that has been masked to remove bright saturated stars and PSF discontinuities from the analysis (see section~\ref{sec:effarea}). We show that the stellar branch, for which \spmod~$\sim 0$, is clearly separated from values for extended objects out to \mauto~$\la 22.5$. 

\begin{figure}
  \includegraphics[width=84mm]{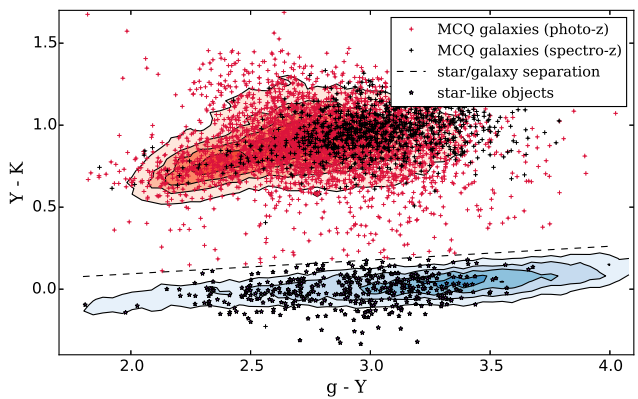}
  \caption{$Y-K$ vs. $g-Y$ colour-colour diagram  used to refine our star/galaxy separation. Blue regions show the locus of high S/N stars ($\textrm{S/N}>10$), whereas red regions indicate all massive quiescent galaxies ($M > 5\times 10^{10}\,M_{\sun}$) with redshifts between 0.2 and 0.6.  MCQ galaxies stands for massive compact quiescent galaxies.}
  \label{fig:starselection}
\end{figure}

For each tile we fit a Gaussian function to the stellar branch, and consider as point-like those objects with a value of \spmod lower than three standard deviations from the mean of the Gaussian (see Figure~\ref{fig:spreadmodel}). We note that the median seeing of the CS82 survey ($0\farcs 6$) is smaller than the median seeing of the SDSS\footnote{\url{http://www.sdss.org/dr12/imaging/other_info/}} ($\sim 1\farcs 43$). \citet{Damjanov2013} showed that some compact galaxies are morphologically identified as stars by SDSS. We have checked that the only object of \citet{Damjanov2013} that lies in Stripe~82 has been well classified as an extended-like object using the procedure described above. However, it is not considered in our analysis due to a stellar mass of $\logten (M_{\star}) = 9.95$, below our mass cut selection.

We use colour information from both SDSS and UKIDSS to remove  misclassified stars from the sample of extended objects. Following \citet{Whitaker2011}, the rest frame colours $u-J$ and $J-K$ are well adapted for a star/galaxy separation. Figure~\ref{fig:starselection} shows the colour-colour diagram $g-Y$ and $Y-K$ of passive massive galaxies with redshifts $0.2 < z < 0.6$ (red regions). The stars selected with the morphological approach are shown in blue contours. Both datasets are clearly separated in this diagram. We define the separation between stars and galaxies by fitting a gaussian function of the projected histogram of the stellar cloud onto the y-axis, allowing for a $2\sigma$ variation. When the $Y$ or $K$~bands were not available, we carried out the separation using $g-J$/$J-K$, $g-H$/$H-K$ and $g-Y$/$Y-H$ colour diagrams, with the caveat that in the two last sets the separation between stars and galaxies is not as clear. For the subset of galaxies with spectroscopic data $-$ identified as galaxies based on their spectral features $-$ we verified that their colours placed them in the ``galaxy region'' of the diagrams. We removed $2.6$\% of the objects of the quiescent catalogue (2,447 out of 94,596) with this extra criterion. 

\subsection{Compactness criteria \label{sec:compact}}

\begin{figure}
  \includegraphics[width=84mm]{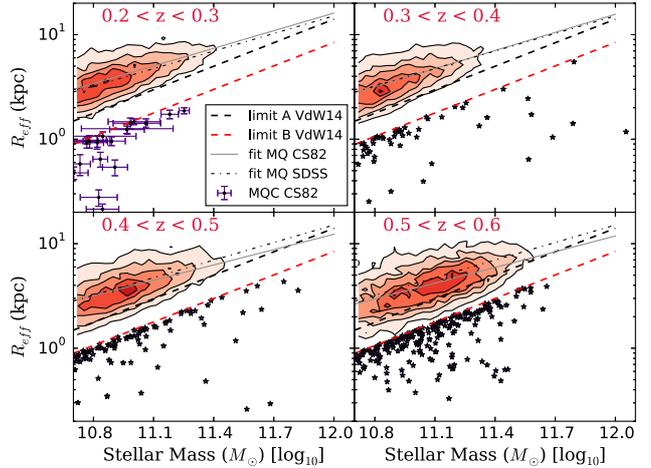}
  \caption{Galaxy size - stellar mass relation in four redshift bins. The quiescent galaxies are shown in red, whereas the most compact ones $-$ according to the strictest criterion of \protect\citet{VanderWel2014} $-$ are shown as indigo stars. The gray line is a linear fit to the distribution of quiescent galaxies in red; to compare we show the fit to the SDSS size distribution by \protect\citet{Shen2003} with the black dot-dashed line. Dashed red and black lines show the `most' and `less' compact criteria by \protect\citet{VanderWel2014}, respectively. For clarity, we only plot the most compact objects according to the \citet{VanderWel2014} criterion and error bars are only shown for the first redshift bin.}
  \label{fig:compactreff}
\end{figure}

As mentioned earlier, we follow the definitions of compactness adopted by \citet{Quilis2013}, \citet{Carollo2013}, \citet{VanderWel2014} and \citet{VanDokkum2015}. While \citet{Carollo2013} and \citet{VanderWel2014} opt for a criterion based on the non-circularized effective radius $\reff$, the other authors use the circularized effective radius $\reffc = \reff \times \sqrt{b/a}$, where b and a are the minor and major axis of the model ellipse containing half of the total flux, respectively. The criteria for a galaxy to be considered as compact by the different authors $-$ and that we integrate into our analysis $-$ are summarized here:
\begin{enumerate}
\item \citet{Quilis2013}: $\reffc < 1.5$\,kpc;
\item \citet{Carollo2013}: we focus on the two lowest size bins of their Figure~4, for which the compact definitions correspond to $\reff < 1.4$\,kpc and $\reff < 2.0$\,kpc, respectively; we refer to these definitions in Table~\ref{tab:numberscompact} and Figure~\ref{fig:density} as the `most' and `less' compact criteria, respectively;
\item \citet{VanderWel2014}: 
\begin{equation}
\reff < A \times \left( M_\star/10^{11}\,M_{\sun} \right)^{0.75} \; ,
\end{equation}
where $A=1.5$\, kpc or $2.5$\,kpc following the most conservative (red dashed line in Figure~\ref{fig:compactreff}) and the loosest (black short-dashed line) criteria of compactness, respectively;
\item \citet{VanDokkum2015}: 
\begin{equation}
\reffc < 2 {\rm ~kpc} \times \left( M_\star/10^{11}\,M_{\sun} \right) \; .
\end{equation}
\end{enumerate}

To characterize the size of our sample we use the effective radius measurement from the de Vaucouleurs fit of CS82 data using \SExtractor and \PSFEx{}. Furthermore, following the same colour selection to separate quiescent and star-forming galaxies as we do in the current analysis, \citet{VanderWel2014} ended up with a quiescent sample for which $80$\% of their S\'ersic index was larger than 2.5. The galaxy size  vs. stellar mass distribution for massive quiescent galaxies is shown in Figure~\ref{fig:compactreff} for four redshift bins in the range of $0.2 < z < 0.6$. The error on the effective radius in kpc ($\reff$) reflects the uncertainty on the redshift and on the measured effective radius in arcseconds ($r_{\textrm{eff}}$); that is, $\delta \reff$\,(kpc)\,$= \left( \delta r_{\textrm{eff}}/r_{\textrm{eff}} + \delta z/z \right) \times \reff$. The error in stellar mass comes from the derivation of the mass with the SYNMAG code. We verified that our size distribution in the lowest redshift bin reproduced the fit performed by \citet{Shen2003} on SDSS data (see Figure~\ref{fig:compactreff}). We confirm the trend of decreasing sizes of early-type galaxies in the past \citep[e.g.,][]{VanderWel2014,Furlong2015}. Although not all compactness criteria are shown here, the resulting samples behave similarly.

\begin{figure}
  \includegraphics[width=\columnwidth]{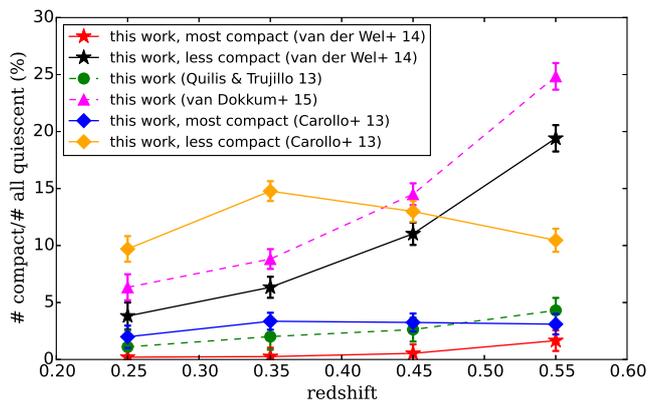}
  \caption{Redshift evolution of the percentage that compact massive quiescent galaxies represent within the general population of massive quiescent galaxies within the redshift range considered in our study (0.2<z<0.6). We show our results considering the different compactness criteria defined in section~\ref{sec:compact} and adopt Poisson error bars.}
  \label{fig:percentage}
\end{figure}

The percentage that compact galaxies represent within the total massive quiescent population is shown as a function of redshift in Figure~\ref{fig:percentage}. Depending on the compactness definition considered, the compact population accounts for merely a few percent up to 25\% of all massive quiescent galaxies.

\section{The catalogue of compact candidates}

\begin{table}
 \caption{Number of massive quiescent galaxies above different minimum masses ($M_{\textrm{min}}$) as defined by the different definitions adopted (C13: \citealt{Carollo2013}; vD15: \citealt{VanDokkum2015}; vdW14: \citealt{VanderWel2014}; QT13: \citealt{Quilis2013}). `Most' and `less' refer to the most and less compact criteria (when it applies) as presented in section~\ref{sec:compact}. The fourth column shows the number of massive compact quiescent candidates selected for each definition, and its respective percentage of galaxies having SDSS spectra.}
 \label{tab:numberscompact}
 \begin{tabular}{cclc}
  \hline
   $M_{\textrm{min}}$ & \# & compact & \# compact \\
   $\logten(M/M_{\sun})$ & quiescent & definition & (with spectra)\\
  \hline
  	$10.5$ & 78,493 & C13 most & 2,381 (0.3\%)\\
  		        & & C13 less & 9,766 (0.6\%)\\
	$10.6$ & 71,408 & vD15 & 9,818 (23\%)\\
	$10.7$ & 62,515 & vdW14 most & 424 (8\%)\\
	            & & vdW14 less & 6,596 (18\%) \\
	$10.9$ & 40,218 & QT13 & 1,103 (2\%) \\
  \hline
 \end{tabular}
\end{table}

We provide in Table~\ref{tab:numberscompact} the number of massive quiescent galaxies and the number of compact ones corresponding to each definition of compactness \citep{Quilis2013, Carollo2013, VanDokkum2015, VanderWel2014}. The percentage of compact candidates that have SDSS spectra is shown. Within this section we only consider the catalogues obtained using the \citet{VanderWel2014} definitions for compactness (most and less conservative ones), as illustrative examples of our analysis; the final results in sections~\ref{sec:completeness} and \ref{sec:results} are shown for all compactness definitions mentioned in section~\ref{sec:compact}. Using the CS82 morphological data, we end up with 424 massive quiescent candidate galaxies for the most compact criterion and 6596 candidates for the less compact one.

\subsection{Morphological properties \label{sec:morphoproperties}}

Two of the authors (AC and EG) visually inspected each of the most and less compact candidates, examining each object, its reconstructed morphological model and the residuals of the model-fitting. They agree on the following classification within an error of $\sim 3$\% among the two and find that:
\begin{enumerate}
\item $\sim 82$\% have a typical elliptical shape and good residuals;
\item $\sim 11$\% have bad residuals due to close neighbours. \SExtractor is currently not able to handle a simultaneous fit of various objects: pixels of the segmentation maps are attributed to only one object, even if receiving signal from two sources; 
\item $\sim 7$\% have bad residuals due to the non adequacy of the de Vaucouleurs shape to fit the galaxy brightness profile. 
\end{enumerate}

We checked that the colours of these three categories were equally distributed in the colour-colour diagram used for the star-forming/quiescent separation. 

\begin{figure}
	    \includegraphics[width=84mm]{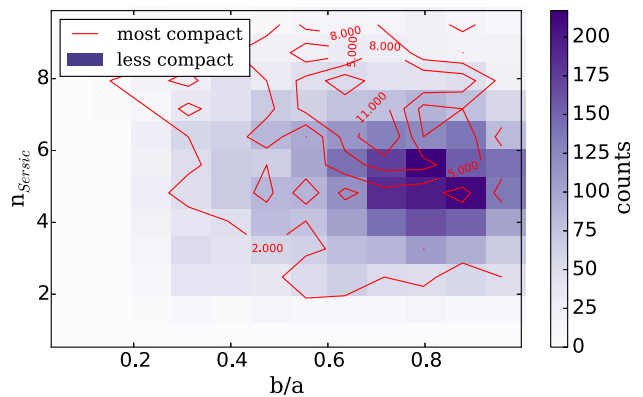}
  	\caption{Distribution of the S\'ersic index as a function of the aspect ratio of the selected compact passive galaxies, extracted from a fit by a general S\'ersic brightness profile. The blue 2D-histogram shows the less compact sample, whereas the red contours represent the most compact candidates. We notice that most compact galaxies are included in the less compact sample.}
  	\label{fig:morphofit}
\end{figure}

In addition to the de Vaucouleurs brightness profile, we fit a general S\'ersic profile to all the objects using \SExtractor. We present in Figure~\ref{fig:morphofit} the S\'ersic index distribution as a function of the aspect ratio (axial ratio of the best-fitting model) for the sample of most and less compact galaxies in red contours and blue colours, respectively (most compact galaxies are included in the less compact sample). Our most and less compact massive quiescent candidates have a high median S\'ersic index ($<n_{\textrm{ser}}>=6.9$ and $<n_{\textrm{ser}}>=5.2$, respectively) characteristic of early-type galaxies ($n_{\textrm{ser}}\geqslant 2.5$); they also present a roundish shape, with a median aspect ratio of $<b/a>=0.66$ and $<b/a>=0.71$, respectively. 

\subsection{Verifying the size estimates \label{sec:sizecheck}}

\begin{figure}
   	\includegraphics[width=84mm]{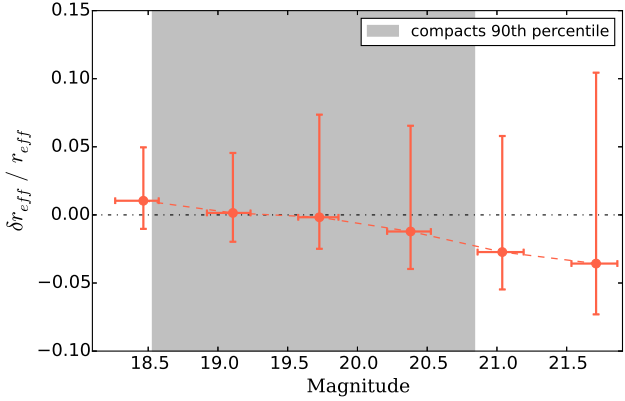}
   	\caption{Relative error on the effective radius as a function of the magnitude of the galaxy. The simulated image contains only pure de Vaucouleurs profiles and is fitted by pure de Vaucouleurs models. Grey area shows the distribution of the 90th percentile of compact candidates selected according to the loosest criterion of \citet{VanderWel2014}.}
  	\label{fig:deVdeltaReffMag}
\end{figure}

\begin{figure}
   	\includegraphics[width=84mm]{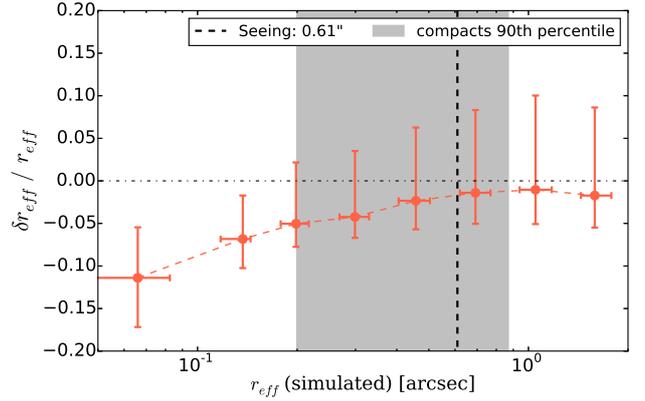}
   	\caption{Relative error on the effective radius as a function of the effective radius of the simulated galaxy. The simulated image contains only pure de Vaucouleurs bulges, and is fitted by pure de Vaucouleurs models. The grey area is defined as in Figure~\ref{fig:deVdeltaReffMag}. The vertical dashed line shows the input seeing of the simulated image.}
  	\label{fig:deVdeltaReff}
\end{figure}

\begin{figure}
   	\includegraphics[width=84mm]{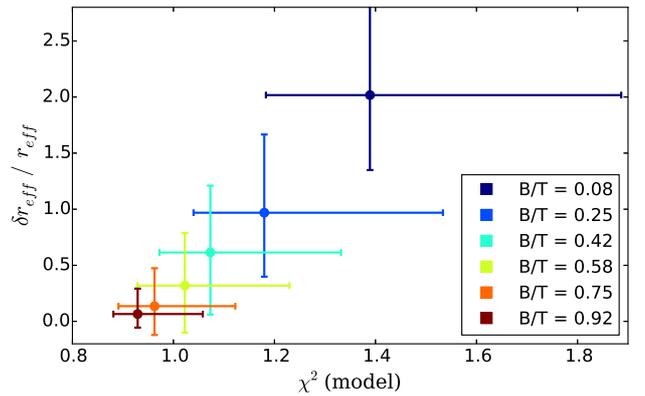}
   	\caption{Relative error on the effective radius as a function of the \SExtractor $\chi^2$ of the fit. The simulated image contains composite profiles described by a sum of an exponential disk and a de Vaucouleurs bulge and is fitted by pure de Vaucouleurs models. The colors vary with the B/T flux ratio.}
  	\label{fig:deVexpdeltaReffchi2}
\end{figure}

\begin{figure}
   	\includegraphics[width=84mm]{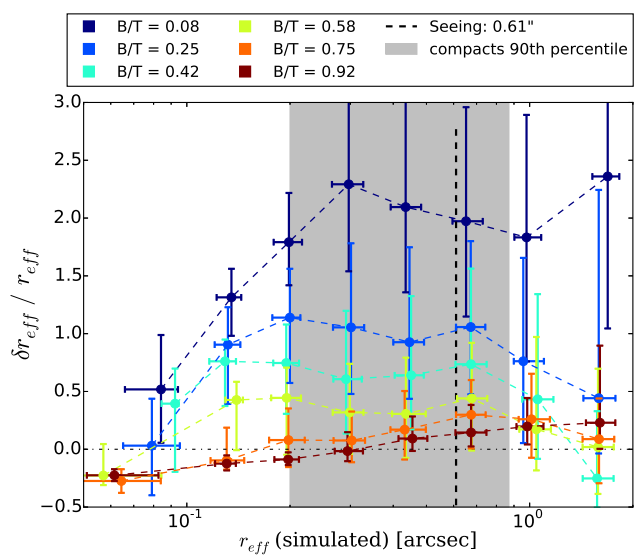}
   	\caption{Relative error on the effective radius as a function of the effective radius of the simulated galaxy. The simulated image contains composite profiles described by a sum of an exponential disk and a de Vaucouleurs bulge and is fitted by pure de Vaucouleurs models. The colors vary with the B/T flux ratio. The Grey area as is defined in Figure~\ref{fig:deVdeltaReffMag}. The vertical dashed line shows the input seeing of the simulated image.}
  	\label{fig:deVexpdeltaReff}
\end{figure}

Our selection of compact galaxies relies on size estimates extracted from ground-based images. There are essentially two main sources of errors: statistic, i.e the ability of the method to recover effective radii generally smaller than the PSF, and systematic, i.e. the uncertainties due to a wrong model assumption. We are indeed using a de Vaucouleurs profile to fit the surface brightness profile of our galaxies. This can induce a systematic error if the model does not properly match the actual galaxy profile \citep[e.g.,][]{Yoon2011, Bernardi2013, Bernardi2017}. The robustness of the \PSFEx and \SExtractor packages in deconvolving the PSF and in estimating the galaxy sizes has to be assessed. We have created simulated images with similar properties to a CS82 coadd image using the \SkyMaker\footnote{\url{http://www.astromatic.net/software/skymaker}} package \citep{Bertin2009}. The gain, exposure time, sky background, seeing and telescope properties were set to be equal to the CS82 ones. The catalogue of morphological properties of a CS82 tile is given as input for \SkyMaker: objects classified as stars in CS82 catalogues are added as PSFs, whereas the properties of all other objects are from our model-fitting. We have created an image containing only stars and pure de Vaucouleurs galaxies, and an image with a large variety of bulge plus disk luminosity profiles modelled by a combination of an exponential and a de Vaucouleurs components. The first set of simulations should allow one to quantify the statistical error (assuming that the model is correct). The second set allows us to estimate the systematic uncertainty arising from using a wrong model to fit the galaxies. The simulated images contain an average of $\sim\,$65,000 objects. We have run \PSFEx and \SExtractor packages on these simulated images using the same pipeline as for real CS82 images assuming a pure de Vaucouleurs profile. The comparison of the sizes from the input catalogues and the recovered ones is shown in Figures~\ref{fig:deVdeltaReffMag} and  \ref{fig:deVdeltaReff} for an image containing only pure de Vaucouleurs profiles, and in Figures~\ref{fig:deVexpdeltaReffchi2} and \ref{fig:deVexpdeltaReff} for an image containing composite brightness models. 
The grey area shows the 90th percentile of compact candidates identified with the loosest criterion of \cite{VanderWel2014}. Points and error bars represent median values and 90th percentiles of the sample defined by $i < 22$ and $r_{\textrm{eff}} < 2$\,arcsec\footnote{$r_{\textrm{eff}}$ is the effective radius $\reff$ (kpc) in arcsec}, containing $\sim\,$9,700 galaxies. We define the relative error on the effective radius as the relative difference between the recovered radius and the radius provided to create the simulated image: $\delta r_{\textrm{eff}}/r_{\textrm{eff}} = (r_{\textrm{eff}} (modelled) - r_{\textrm{eff}} (simulated))/r_{\textrm{eff}} (simulated)$. 
For the image created with pure de Vaucouleurs profiles and fitted by pure de Vaucouleurs models, we find that the relative error depends only weakly on the magnitude of the objects (see Figure~\ref{fig:deVdeltaReffMag}). We observe a systematic negative offset varying from $5$ to $1$\% for smaller (0,2\,arcsec) to larger galaxies (0,9\,arcsec), however compatible with 0 (see Figure~\ref{fig:deVdeltaReff}). This offset suggests that we are underestimating the size of the galaxies at the $\sim 2-5$\% level if the galaxy is effectively a pure bulge. 
On the other hand if the galaxy is a sum of two components, a bulge plus a disk, the quality of the fit by a pure de Vaucouleurs profile is strongly dependent on the bulge-to-total (B/T) flux ratio. The $\chi^2$ of the \SExtractor model is not normalized to 1, but indicates better fit for $\chi^2 \sim 0.85$. Figure~\ref{fig:deVexpdeltaReffchi2} shows the increase of the $\chi^2$ with the decrease of the B/T ratio. The relative error on the effective radius is directly related to the B/T ratio: for $\textrm{B/T}<0.5$, the relative error becomes larger than $50$\%. Figure~\ref{fig:deVexpdeltaReff} illustrates the dependence of $\delta r_{\textrm{eff}}/r_{\textrm{eff}}$ on the input size of the galaxy. The effective radius is overestimated whatever the B/T ratio for the range of compact galaxy sizes (except for pure small bulges). For $\textrm{B/T}>0.75$, the error reaches $\sim 30$\%. It means that if a compact galaxy is not a pure de Vaucouleurs bulge and has another component, our methodology tends to overestimate the size. As discussed in section~\ref{sec:conclusion}, this will impact our results in the sense that we might lose a fraction of the compact candidates. If the error on the galaxy size estimate is systematic at the level of 5\%, with all other equal parameters, this would lead to an increase of $33$\% and $28$\% in the number density of the most and less compact galaxies, respectively. Therefore, if anything, the measured number densities are underestimated. Accounting for this error makes our conclusions even stronger. 

\begin{figure}
    \includegraphics[width=84mm]{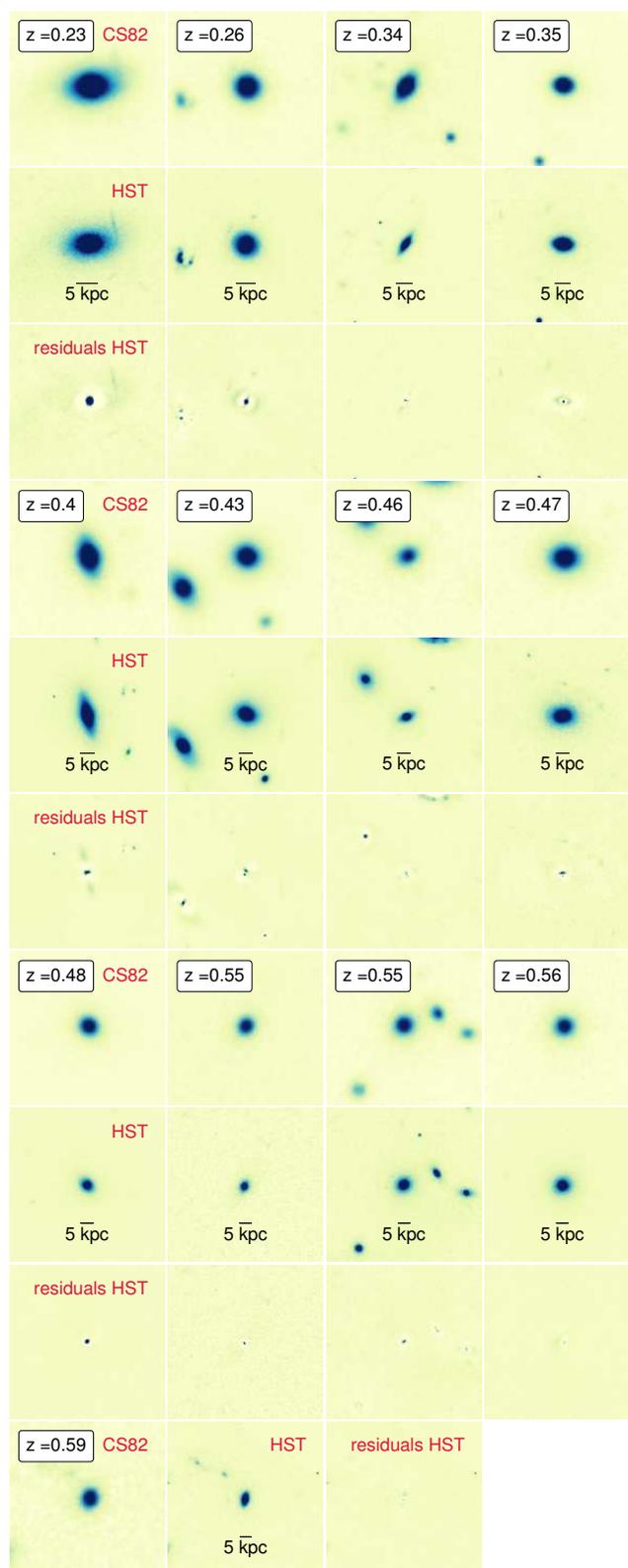}
  \caption{Compact massive quiescent galaxies that have been observed by both CS82 (first, fourth, seventh and tenth lines) and HST/ACS (second, fifth, eighth and tenth lines) in $i$-band. The images are centred on the galaxy of interest. The residuals of the fitting of the HST images by a de Vaucouleurs brightness profile are shown (third, sixth, ninth and tenth lines). The cutouts are north-oriented, with a side size of $10\arcsec$. An inset bar indicates the size of 5\,kpc at the redshift of the galaxies.}
  \label{fig:hstimages}
\end{figure}
We also checked the reliability of the PSF deconvolution by the \PSFEx package by comparing the results with space based images of higher quality. We collected HST images taken with the Advanced Camera for Surveys (ACS) Wide field Channel (WFC) in the $i$-bands (nominally the F814W and F775W filters) from the Hubble Legacy Archive.\footnote{\url{http://hla.stsci.edu}} Thirteen of our massive compact candidates within the less conservative sample have been covered by HST. The images come from different observing programs and their exposure time varies between $1{,}000$ and $7{,}866$~seconds, corresponding respectively to depths of $\sim 23.5$ and $24.5$ in $i$-band for extended sources. We estimated the morphological parameters by running \SExtractor and \PSFEx on each HST image. All compact candidates are indeed extended objects (i.e., not stars), as shown in Figure~\ref{fig:hstimages}. The redshifts indicated in boxes are spectroscopic (SDSS/BOSS) for four of them (the ones with $z=0.4, 0.47, 0.56$ and the second in the row with $z=0.55$), and photometric for the remaining nine.

\begin{figure}
   	\includegraphics[width=84mm]{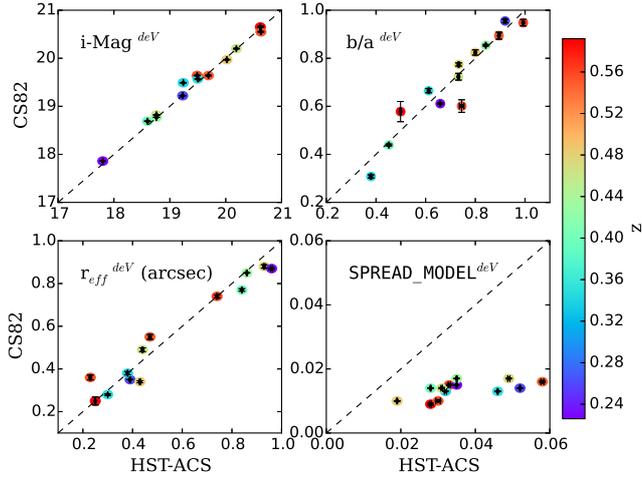}
   	\caption{Comparison of the morphological and photometric properties of the compact galaxies that have been observed both by CS82 and HST. The parameters were derived by fitting a de Vaucouleurs brightness profile using \SExtractor and \PSFEx: model magnitude (upper left), aspect ratio $b/a$ (upper right), effective radius $r_\textrm{eff}$ (lower left) and the \spmod parameter (lower right). The colour scale indicates the redshift.}
  	\label{fig:hstcs82}
\end{figure}

We compare the morphological outputs from \SExtractor obtained by fitting a de Vaucouleurs profile on both sets of images (see Figure~\ref{fig:hstcs82}). The photometry and the morphology extracted from both CS82 and HST/ACS data are in very good agreement. We report a mean squared difference of $0.10$ in magnitude and of $0.4$\,kpc for the effective radius. We do not identify clear systematics due to the larger PSF of CS82 images and we verify the robustness of the PSF deconvolution by the \PSFEx code. The \spmod parameters derived from the HST images are larger than the CS82 as the galaxies are clearly resolved as extended sources. 

\section{Number density evolution}

\subsection{Effective area \label{sec:effarea}}

\begin{figure}
  \begin{center}
    \includegraphics[trim={15 3 15 3},clip,width=84mm]{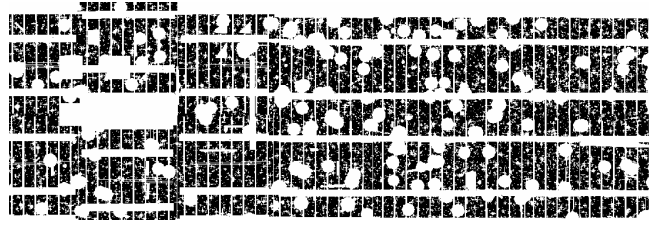}
  \end{center}
  \caption{Combined masks from CS82 and UKIDSS in a portion of Stripe~82 for the estimate of the effective area.}
  \label{fig:masks}
\end{figure}

One characteristic that sets this study aside from other works \citep[e.g.,][]{Valentinuzzi2010a, Valentinuzzi2010b, Poggianti2013a, Damjanov2014, Tortora2016} is the uniform coverage of a large contiguous region of the sky, without pre-selection based on the environment. To calculate the evolution of the density of compact galaxies with redshift, we first estimate the area covered by the CS82 survey. We combined the masks of UKIDSS and CS82 using the \textsc{WeightWatcher}%
\footnote{\url{http://www.astromatic.net/software/weightwatcher}} and \textsc{SWarp}\footnote{\url{http://www.astromatic.net/software/swarp}} packages. \textsc{WeightWatcher} is designed to combine weight maps, flag maps and polygons, whereas \textsc{SWarp} re-samples and coadds FITS images. The CS82 and UKIDSS masks were produced to remove bright stars, cosmic rays and artefacts. Considering that CS82 images are built from four single exposures, dithered to fill the gap between CCD chips and that our method is sensitive to PSF discontinuities that appear in these regions, we omit from our search the $\sim 33$\% of the total area corresponding to these interstitial regions. We show a portion of the total mask in Figure~\ref{fig:masks}: the horizontal and vertical lines correspond to the inter-CCD regions. We have measured an effective area of $A_{\textrm{eff}} = 82.6 \pm 7.3$\,deg$^2$. 

\subsection{Completeness \label{sec:completeness}}

\begin{figure}
	    \includegraphics[width=84mm]{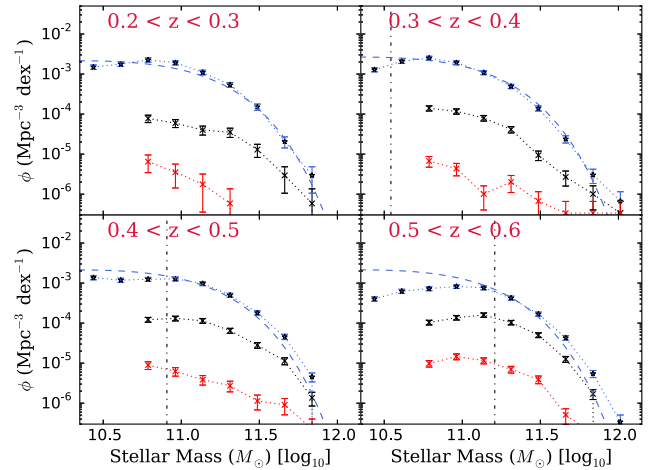}
 	 \caption{Galaxy stellar mass functions (GSMF) in four bins of redshift between $z=0.2$ and $z=0.6$. Blue upper point represent the quiescent selections, lower red and middle black points show the most and less compact massive candidates obtained adopting the \citet{VanderWel2014} criteria, respectively. Vertical magenta dash-dotted lines show the limiting mass above which the quiescent sample is complete. Blue dashed lines are double Schechter functions fitted to quiescent GSMF data, whose parameters are taken from \citet{Ilbert2013}. In the case of the first panel, the vertical line lies at smaller values than the stellar mass range.}
  	\label{fig:massfunction}
\end{figure}

Our sample of compact massive quiescent galaxies suffers incompleteness at the low mass end due to the magnitude limit of the NIR data (optical data are deeper). We derive a corresponding 80\% completeness magnitude limit in $i$-band of $\sim 20.5$ for extended sources by looking at the inflection in the number counts in the $i$-band. This magnitude limit in $i$-band corresponds to an increasing limiting stellar mass at increasing redshifts, above which we consider that the sample of quiescent galaxies is complete. The limiting stellar mass is defined as in \citet{Pozzetti2010}: for each bin of redshift in $z=[0.2,0.3,0.4,0.5,0.6]$ we compute the upper envelope of the limiting mass distribution for 95\% completeness and find $10.2$, $10.5$, $10.9$ and $11.2$ (in $\logten M_\star /M_{\sun}$), respectively. These values are represented by vertical lines in Figure~\ref{fig:massfunction} and summarized in Table~\ref{tab:lossfactor}.

We compute the galaxy stellar mass function (GSMF) following the $V_{\rm max}$ method \citep{Schmidt1968, Baldry2008}:
\begin{equation}
\Phi_{\logten M} = \frac{1}{\Delta \logten M} \sum\limits_{j} \frac{1}{V_{\rm max, j}} \, ,
\end{equation}
where $\Delta \logten M$ is the mass bin in logarithmic scale. $V_{\rm max, j}$ is the comoving volume over which the $j$th galaxy could be observed and is computed given the redshift of the galaxy and the limiting magnitude of the survey ($21.0$ in $i$-band magnitude at $50$\% completeness). Our quiescent GSMF is shown in Figure~\ref{fig:massfunction} as blue stars in different redshift bins. To estimate how much of the passive galaxy population we are losing at higher redshift towards the lower masses, we assume that the shape of the GSMF does not vary significantly between $z=0.2$ and $z=0.6$, and that we can simply renormalise the double Schechter function provided by \citet{Ilbert2013}. This renormalization is shown as blue dashed lines in Figure~\ref{fig:massfunction}, where we have only included in the fit the data points that are above the limiting mass. We define the completeness factor as the ratio of the number of detected galaxies over the number of expected galaxies according to the Schechter function. Completeness factors computed for different minimum stellar masses ($10^{10.5}$, $10^{10.6}$, $10^{10.7}$ and $10^{10.9}\,M_{\sun}$) are summarized in Table~\ref{tab:lossfactor}. We find that we are complete towards lower redshifts where completeness factors are close or equal to one for all the chosen minimum stellar masses of our samples. We miss less massive galaxies in number counts at higher redshifts: for the bin $0.5<z<0.6$ our selection of quiescent galaxies is complete at 57\% above $10^{10.5}\,M_{\sun}$ and at 80\% above $10^{10.9}\,M_{\sun}$.

\begin{table}
 \caption{Limiting stellar mass, for each redshift bin, above which the sample of quiescent galaxies is complete, and factor of completeness for subsets of galaxies above a given minimum stellar mass ($10^{10.5}$, $10^{10.6}$, $10^{10.7}$ and $10^{10.9}\,M_{\sun}$).}
 \label{tab:lossfactor}
 \begin{tabular}{lccccc}
  \hline
  z bin & limiting mass & \multicolumn{4}{c}{completeness factor} \\ 
   & $\logten(M_\star /M_{\sun})$ & \multicolumn{4}{c}{for $\logten(M_\star /M_{\sun})$}\\
  & &  $> 10.5$ & $> 10.6$ & $> 10.7$ & $>10.9$ \\
  \hline
  0.2-0.3 & 10.18 & 1 & 1 & 0.99 & 0.97 \\  
  0.3-0.4 & 10.54 & 1 & 0.99 & 0.97 & 0.96 \\ 
  0.4-0.5 & 10.91 & 0.73 & 0.84 & 0.95 & 1 \\
  0.5-0.6 & 11.21 & 0.57 & 0.59 & 0.63 & 0.80 \\
  \hline
 \end{tabular}
\end{table}

We take into account these completeness factors in the estimate of the number densities that are shown on Figures~\ref{fig:density} and~\ref{fig:densitysimulations} where the raw number count of compact quiescent galaxies above a given stellar mass in a given redshift bin is divided by the appropriate completeness factor taken from Table~\ref{tab:lossfactor}. We have calculated the influence of the slope of the GSMF at the low mass end on the number density of compact massive galaxies by artificially changing the limiting stellar mass within the stellar mass median error ($\mathrm{d} \logten(M_\star /M_{\sun}) = 0.09$). The value of the resulting error on the number density depends on the chosen compactness definition and on the redshift bin, but it is on the order of $\sim 10$\%. This error is included in Figures~\ref{fig:density} and~\ref{fig:densitysimulations}.

For illustration purposes, we also compute the compact quiescent GSMF for the samples following the \citet{VanderWel2014} compactness definition. Less (black crosses) and most (red crosses) compact candidates are shown in Figure~\ref{fig:massfunction}. They follow the global behaviour of quiescent galaxies. In the lower redshift bin ($0.2 < z < 0.3$) the knee of the GSMF is only visible for the less compact sample.

\subsection{Results \label{sec:results}}

\begin{figure*}
	    \includegraphics[width=\textwidth]{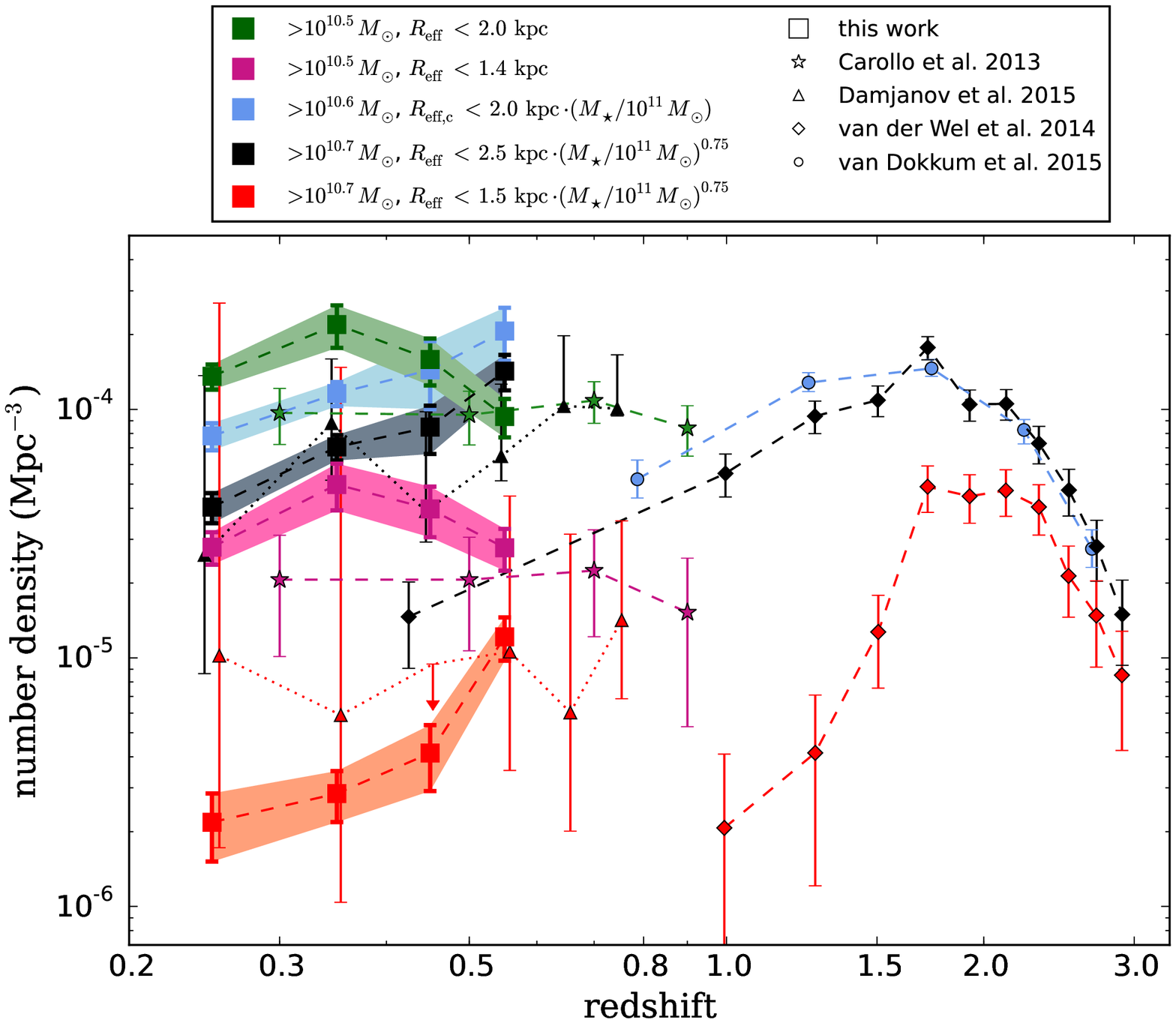}
	  \caption{Evolution of the number density of quiescent compact massive galaxies down to redshift 3. Different colours indicate different definitions of compactness: red and black for the most and less conservative criteria of \citet{VanderWel2014}, respectively (see section~\ref{sec:compact}), blue refers to the one adopted by \citet{VanDokkum2015}, pink and green to the most and less conservative criteria of \citet{Carollo2013}, respectively. Densities obtained with CS82 data are shown with large squares (this work), diamonds represent \citet{VanderWel2014} data, circles \citet{VanDokkum2015} and stars \citet{Carollo2013} ones. At intermediate redshifts we also plot as triangles the results of \citet{Damjanov2015} based on the COSMOS data, following the same colour code.}
 	 \label{fig:density}
\end{figure*}

In Figure~\ref{fig:density} we compare our results for the variation of the number density of massive quiescent compact galaxies in the Stripe~82 over the redshift range $0.2 < z < 0.6$ to their counterparts at higher redshifts in the literature, using the same definitions of compactness \citep{Carollo2013, VanderWel2014, VanDokkum2015}. We observe that the density of massive compact quiescent galaxies at intermediate redshifts decreases towards lower redshifts for two of the compactness criteria \citep{VanderWel2014, VanDokkum2015}, while it remains approximately constant following the criteria of \citet{Carollo2013}. We find that the number density of the most compact samples of \citet{VanderWel2014} and \citet{Carollo2013}  are 1.3 dex and 0.8 smaller than the number density of their corresponding less compact selections, respectively. Both the minimal mass of the sample and the compactness definition have an influence on the behaviour of the derived number density of compact massive galaxies. Within the error bars, we confirm the trend observed by \citet{Carollo2013} from redshifts $z = 1.0$: the number density of compact galaxies with a mass larger than $10^{10}\,M_{\sun}$ is roughly stable since redshift $z\sim 0.8$. Our data does not connect easily to other higher redshift data: \citet{VanderWel2014} and  \citet{VanDokkum2015} observe the beginning of a decrease at redshift $z\sim 1.5$ that leads to values at intermediate redshifts that are lower by a factor $\sim 5-6$ than our observations. However, \citet{Damjanov2015} observe the same trend as us at intermediate redshifts working on the COSMOS field and applying the compactness criteria of \citet{VanderWel2014}, albeit with larger error bars. The gap is likely due to the limited volume of CANDELS at intermediate redshifts or to a bias in the sample selection and definition. We observe the same trend as \citet{Cassata2013}: the number density of smaller early-type galaxies evolves more rapidly than that of larger ones. We do not compare directly our density with their result as they adopt a minimal stellar mass of $10^{10}\,M_{\sun}$, that we consider to be too low in the context of our paper.

\begin{figure}
	    \includegraphics[width=84mm]{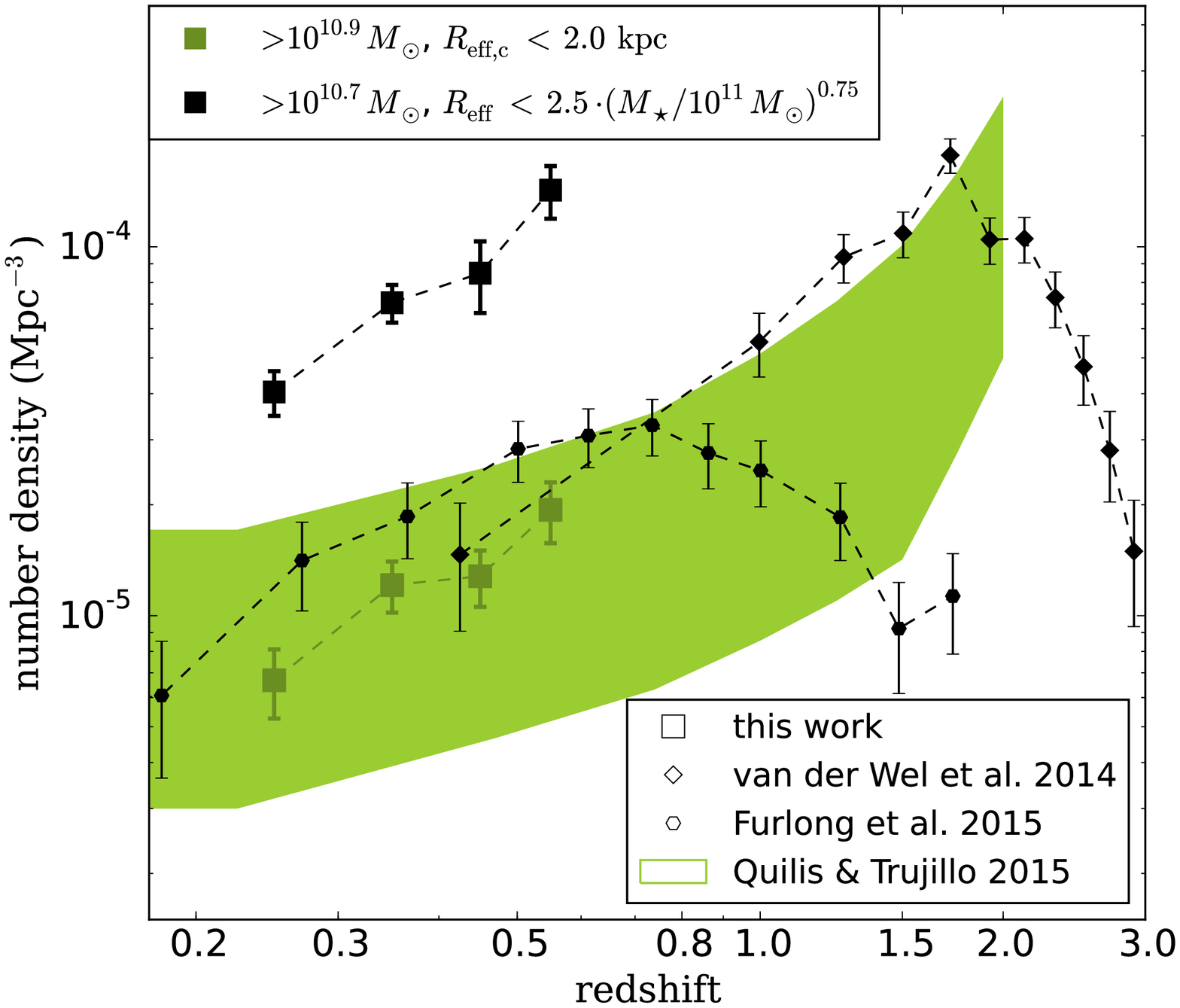}
	  \caption{Evolution over cosmic times of the number density of compact massive quiescent galaxies. Our results are compared with simulations. Large squares present this work, following two different compactness definitions: in black the loosest definition of \citet{VanderWel2014} and in green the one of \citet{Quilis2013}. The green area shows the expectations of \citet{Quilis2013} from semianalytical models. Black hexagons show the prediction of \citet{Furlong2015} in the context of the EAGLE hydrodynamical simulation. Black diamonds show the less compact criterion of \citet{VanderWel2014}.}
 	 \label{fig:densitysimulations}
\end{figure}

Current N-body simulations have provided some clues to understand the evolution of the population of compact massive quiescent galaxies with cosmic times. \citet{Furlong2015} have applied the less conservative criterion of compactness of \citet{VanderWel2014} on the EAGLE simulation. The gravitational force softening length (i.e. the spatial resolution) of the largest EAGLE simulation used in their analysis is $\epsilon = 0.70$~kpc. They test the convergence of their results by comparing runs of various resolutions. Below the resolution of the simulations, subgrid models are applied. The gravitational softening is smaller than the strongest criteria of \citet{VanderWel2014}, making them adequate for comparison. They find that the number density of compact massive quiescent galaxies increases for decreasing redshifts until $z\sim 0.7$, then declines for $z<0.8$. At high redshifts, the discrepancy between their data and the observed density by \citet{VanderWel2014} is likely due to the limited box size of the simulation. However, we emphasize that the comparison with observations is not that simple, because the determination of the effective radius and the stellar mass are done in different ways. We show the comparison of our measurements at intermediate redshifts with the EAGLE measurements in Figure~\ref{fig:densitysimulations}. At intermediate redshifts, they expect a continuous decrease of the number density of massive compact galaxies that we do observe in our study but with an offset of $\sim 0.7$\,dex. \citet{Quilis2013} used semianalytical models based on the Millenium simulation to calculate the expected fraction of massive compact galaxies that remain almost untouched since redshifts $z>2$, i.e. that evolve in stellar mass by less than 30\%. Applying similar cuts in mass and circularized effective radii (see sections~\ref{sec:samplesel} and~\ref{sec:compact}), we obtain number densities that are within the error bars of the expected value of \citet{Quilis2013} and that follow the same trend. Finally \citet{Wellons2016} also predict that the number density of compact massive quiescent galaxies should decrease in the local universe; although they attribute this to the processes that galaxies undergo during their evolution, they provide no further quantification of these conclusions. 

We investigate the parameters of the compact massive galaxy definition that lead to different behaviours of the number density at intermediate redshifts. We identify that the increase of the number density is strongly related to the lower limit in mass of the sample of massive galaxies. Applying the minimal masses of \citet{VanderWel2014} and \citet{Quilis2013} as defined in section~\ref{sec:samplesel} associated to the compact criterion of \citet{Carollo2013}, we do observe an increase of the number density of compact massive quiescent galaxies at intermediate redshifts. This is confirmed by \citet{Carollo2013} concerning their most massive sample of compact quiescent galaxies.

\section{Discussion and concluding remarks}
\label{sec:conclusion}

Hydrodynamical simulations suggest that the compact passive massive galaxy population is continuously evolving. According to \citet{Furlong2015} and \citet{Wellons2016} the main size growth mechanisms of passive galaxies between $z>1.5$ and $z=0$ are acquisition of ex-situ stars through dry-merger events and renewed star formation events also triggered by mergers. Thus very few will have been left untouched with cosmic time, making untouched relics very rare. Studying the number density evolution of compact massive relics is unlikely to reflect the evolution of individual galaxies, but instead gives indications about the frequency of the merging processes that this population encounters over cosmic times. Comparing our results with simulations, we thus confirm the stochastic behaviour of the minor merging processes. We observe a constant decrease of the number density of this galaxy population at intermediate redshifts, and notice that the normalization and the behaviour of the number density is clearly sensitive to the definitions adopted to characterize compact relics. Our results are in agreement with the conclusions of \citet{Carollo2013}: we indeed find that newly quenched galaxies may have typical sizes larger than high redshift ones to explain the progressive disappearance of compact massive galaxies. We however observe an individual evolution of this relic population, that is confirmed by adopting minimal masses larger than $M_\star > 5\times 10^{10}\,M_{\sun}$ instead of $M_\star > 3\times 10^{10}\,M_{\sun}$ as in \citet{Carollo2013}. The number density of compact quiescent galaxies is particularly sensitive to the mass interval considered. 

The population of compact galaxies at intermediate redshifts is scarce and therefore requires a large survey area to have enough statistics. Stripe 82 data complies with many crucial points in this context and allow us to reduce significantly the error bars on the compact relic number density. We note an offset between our observations and the \citet{Furlong2015} predictions (see Figure~\ref{fig:densitysimulations}) and attribute it to potential environmental effects, considering that the volume probed by hydrodynamical simulations is smaller than ours. This effect known as cosmic variance has an influence on galaxy population properties. \citet{Moster2011} show that for the COSMOS, EGS and GOODS fields, we should expect a cosmic variance of $\sim 0.2$, $\sim 0.25$ and $\sim 0.35$ for massive galaxies at intermediate redshifts, respectively. Moreover, as underlined by \citet{Stringer2015}, \citet{Wellons2016} and \citet{PeraltaDeArriba2016}, the environment of compact galaxies plays a critical role with respect to their potential survival. We note that although these studies all agree on this matter, they have conflicting conclusions on the specifics: the first study alleges that isolated galaxies are more likely to be protected from merger events, whereas the other two studies point to the dense central regions of galaxy clusters as the most likely places to find relics. In a future paper we envisage exploring the impact of environment on the compact galaxy population in the context of the Stripe 82 survey, with a  sky coverage that does not suffer from pre-selection based on the environment.

Despite the fact that we have based our analysis on ground based images of excellent quality, we cannot exclude that there is possible contamination in our sample coming from stars and from inaccurate morphological parameters. Size estimates might be a source of systematics, in particular for galaxies that have close neighbours; these represent $\sim 11$\% of the total sample. The fit of the surface brightness profile by a de Vaucouleurs profile with the \PSFEx and \SExtractor packages results in size overestimates for galaxies that are not adequately described by a pure bulge. This translates into an underestimate of the number densities of compact galaxies (see section~\ref{sec:sizecheck}). The population of compact quiescent galaxies that we have identified is therefore conservative and our conclusions will not be affected but strengthened by this systematic effect.

\bigskip

In this paper, we have identified a population of quiescent massive compact galaxies at intermediate redshifts, making use of the exceptional multiwavelength coverage of the equatorial region called Stripe~82. Morphological parameters were derived running \SExtractor and \PSFEx codes on CFHT/Megacam deep $i$-band images from CS82. We apply different definitions of compactness to compare our results to previous studies. We find that:

\begin{enumerate}
\item There is a strong dependence of the absolute number density of compact massive galaxies with the adopted compactness definition. It varies e.g. by a factor of $\sim 80$ between the \citet{Carollo2013} and the strictest \citet{VanderWel2014} definitions. This variation is significantly larger than the errors on the number density. 

\item The number density  of compact massive galaxies evolves relatively slowly at intermediate redshifts. It decreases with cosmic time by a factor of  $\sim 5$ between $z=0.6$ and $z=0.2$ when adopting the \citet{VanderWel2014} or \citet{VanDokkum2015} definitions and remains constant within error bars according to the compactness definition of \citet{Carollo2013}. We note that the evolution of the number density with redshifts is significantly smaller than the absolute variation due to the adopted compactness definition.

\item A significant offset in number density is observed between our measurements at intermediate redshifts and previous works. We systematically find larger number densities by a factor of $\sim 5$ compared to \citealt{VanderWel2014} and \citealt{VanDokkum2015} at $z\sim 0.6$. Cosmic variance might explain this difference as our volume at that redshift is $\sim 330$ times larger than the CANDELS one. Our measurements at $z=0.6$ are roughly compatible with the number densities obtained at redshifts 1.5-2 by \citealt{VanderWel2014} and by \citealt{VanDokkum2015}. This lack of evolution suggests that most of the size evolution observed in these populations is due to progenitor bias. Only the abundance of extreme compact galaxies (the most compact galaxies of \citealt{VanderWel2014}) seem to have dropped by a factor of 20 since $z=2$. This is likely due to the disappearance of very compact progenitors below $z<2$ and to the global size growth of early type galaxies over cosmic times. This confirms the stochastic behaviour of merging processes observed by hydrodynamical and cosmological simulations.

\end{enumerate}

\section*{Acknowledgements}
We thank our anonymous referee for useful comments that improved this paper. AC is supported by the Brazilian Science Without Borders program, managed by the Coordena\c{c}\~ao de Aperfei\c{c}oamento de Pessoal de N\'ivel Superior (CAPES) fundation, and the Conselho Nacional de Desenvolvimento Cient\'ifico e Tecnol\'ogico (CNPq) agency. Fora Temer (FT). KMD and TSG thank the support of the Productivity in Research Grant of the Brazilian National Council for Scientific and Technological Development (CNPq). MM is partially supported by CNPq (grant 312353/2015-4) and FAPERJ (grant E-26/110.516/2-2012), FT. TE is supported by the Deutsche Forschungsgemeinschaft in the framework of the TR33 `The Dark Universe'. HHi acknowledges support from the DFG under Emmy Noether grant Hi 1495/2-1. HYS acknowledges the support from Marie-Curie International Incoming Fellowship (FP7-PEOPLE-2012-IIF/327561) and NSFC of China under grants 11103011. CBG acknowledges financial support from PRIN-INAF 2014 1.05.01.94.02. We thank I. Trujillo, E. Bertin and the LASEX\footnote{\url{http://dgp.cnpq.br/dgp/espelhogrupo/5167044310442074}, Laborat\'orio de Astrof\'isica Extragal\'actica do Observat\'orio do Valongo} members for fruitful discussions. This work is based on observations obtained with MegaPrime/MegaCam, a joint project of CFHT and CEA/DAPNIA, at the Canada-France-Hawaii Telescope (CFHT), which is operated by the National Research Council (NRC) of Canada, the Institut National des Sciences de l'Univers of the Centre National de la Recherche Scientifique (CNRS) of France, and the University of Hawaii. The Brazilian partnership on CFHT is managed by the Laborat\'orio Nacional de Astrof\'isica (LNA). We thank the support of the Laborat\'orio Interinstitucional de e-Astronomia (LIneA). We thank the CFHTLenS team.


\bibliographystyle{mnras}
\bibliography{compactscs82}

\begin{thebibliography}{}
\makeatletter
\relax
\def\mn@urlcharsother{\let\do\@makeother \do\$\do\&\do\#\do\^\do\_\do\%\do\~}
\def\mn@doi{\begingroup\mn@urlcharsother \@ifnextchar [ {\mn@doi@}
  {\mn@doi@[]}}
\def\mn@doi@[#1]#2{\def\@tempa{#1}\ifx\@tempa\@empty \href
  {http://dx.doi.org/#2} {doi:#2}\else \href {http://dx.doi.org/#2} {#1}\fi
  \endgroup}
\def\mn@eprint#1#2{\mn@eprint@#1:#2::\@nil}
\def\mn@eprint@arXiv#1{\href {http://arxiv.org/abs/#1} {{\tt arXiv:#1}}}
\def\mn@eprint@dblp#1{\href {http://dblp.uni-trier.de/rec/bibtex/#1.xml}
  {dblp:#1}}
\def\mn@eprint@#1:#2:#3:#4\@nil{\def\@tempa {#1}\def\@tempb {#2}\def\@tempc
  {#3}\ifx \@tempc \@empty \let \@tempc \@tempb \let \@tempb \@tempa \fi \ifx
  \@tempb \@empty \def\@tempb {arXiv}\fi \@ifundefined
  {mn@eprint@\@tempb}{\@tempb:\@tempc}{\expandafter \expandafter \csname
  mn@eprint@\@tempb\endcsname \expandafter{\@tempc}}}

\bibitem[\protect\citeauthoryear{{Abazajian} et~al.,}{{Abazajian}
  et~al.}{2009}]{Abazajian2009}
{Abazajian} K.~N.,  et~al., 2009, \mn@doi [\apjs]
  {10.1088/0067-0049/182/2/543}, \href
  {http://adsabs.harvard.edu/abs/2009ApJS..182..543A} {182, 543}

\bibitem[\protect\citeauthoryear{{Aihara} et~al.,}{{Aihara}
  et~al.}{2017}]{Aihara2017}
{Aihara} H.,  et~al., 2017, preprint, \href
  {http://adsabs.harvard.edu/abs/2017arXiv170208449A} {} (\mn@eprint {arXiv}
  {1702.08449})

\bibitem[\protect\citeauthoryear{{Annis} et~al.,}{{Annis}
  et~al.}{2014}]{Annis2014}
{Annis} J.,  et~al., 2014, \mn@doi [\apj] {10.1088/0004-637X/794/2/120}, \href
  {http://adsabs.harvard.edu/abs/2014ApJ...794..120A} {794, 120}

\bibitem[\protect\citeauthoryear{{Annunziatella}, {Mercurio}, {Brescia},
  {Cavuoti}  \& {Longo}}{{Annunziatella} et~al.}{2013}]{Annunziatella2013}
{Annunziatella} M.,  {Mercurio} A.,  {Brescia} M.,  {Cavuoti} S.,   {Longo} G.,
   2013, \mn@doi [\pasp] {10.1086/669333}, \href
  {http://adsabs.harvard.edu/abs/2013PASP..125...68A} {125, 68}

\bibitem[\protect\citeauthoryear{{Baldry}, {Glazebrook}, {Brinkmann},
  {Ivezi{\'c}}, {Lupton}, {Nichol}  \& {Szalay}}{{Baldry}
  et~al.}{2004}]{Baldry2004}
{Baldry} I.~K.,  {Glazebrook} K.,  {Brinkmann} J.,  {Ivezi{\'c}} {\v Z}.,
  {Lupton} R.~H.,  {Nichol} R.~C.,   {Szalay} A.~S.,  2004, \mn@doi [\apj]
  {10.1086/380092}, \href {http://adsabs.harvard.edu/abs/2004ApJ...600..681B}
  {600, 681}

\bibitem[\protect\citeauthoryear{{Baldry}, {Glazebrook}  \& {Driver}}{{Baldry}
  et~al.}{2008}]{Baldry2008}
{Baldry} I.~K.,  {Glazebrook} K.,   {Driver} S.~P.,  2008, \mn@doi [\mnras]
  {10.1111/j.1365-2966.2008.13348.x}, \href
  {http://adsabs.harvard.edu/abs/2008MNRAS.388..945B} {388, 945}

\bibitem[\protect\citeauthoryear{{Barro} et~al.,}{{Barro}
  et~al.}{2013}]{Barro2013}
{Barro} G.,  et~al., 2013, \mn@doi [\apj] {10.1088/0004-637X/765/2/104}, \href
  {http://adsabs.harvard.edu/abs/2013ApJ...765..104B} {765, 104}

\bibitem[\protect\citeauthoryear{{Bernardi}, {Meert}, {Sheth}, {Vikram},
  {Huertas-Company}, {Mei}  \& {Shankar}}{{Bernardi}
  et~al.}{2013}]{Bernardi2013}
{Bernardi} M.,  {Meert} A.,  {Sheth} R.~K.,  {Vikram} V.,  {Huertas-Company}
  M.,  {Mei} S.,   {Shankar} F.,  2013, \mn@doi [\mnras]
  {10.1093/mnras/stt1607}, \href
  {http://adsabs.harvard.edu/abs/2013MNRAS.436..697B} {436, 697}

\bibitem[\protect\citeauthoryear{{Bernardi}, {Fischer}, {Sheth}, {Meert},
  {Huertas-Company}, {Shankar}  \& {Vikram}}{{Bernardi}
  et~al.}{2017}]{Bernardi2017}
{Bernardi} M.,  {Fischer} J.-L.,  {Sheth} R.~K.,  {Meert} A.,
  {Huertas-Company} M.,  {Shankar} F.,   {Vikram} V.,  2017, preprint, \href
  {http://adsabs.harvard.edu/abs/2017arXiv170208527B} {} (\mn@eprint {arXiv}
  {1702.08527})

\bibitem[\protect\citeauthoryear{{Bertin}}{{Bertin}}{2009}]{Bertin2009}
{Bertin} E.,  2009, \memsai, \href
  {http://adsabs.harvard.edu/abs/2009MmSAI..80..422B} {80, 422}

\bibitem[\protect\citeauthoryear{{Bertin}}{{Bertin}}{2011}]{Bertin2011}
{Bertin} E.,  2011, in {Evans} I.~N.,  {Accomazzi} A.,  {Mink} D.~J.,   {Rots}
  A.~H.,  eds,  Astronomical Society of the Pacific Conference Series Vol. 442,
  Astronomical Data Analysis Software and Systems XX. p.~435

\bibitem[\protect\citeauthoryear{{Bertin} \& {Arnouts}}{{Bertin} \&
  {Arnouts}}{1996}]{Bertin1996}
{Bertin} E.,  {Arnouts} S.,  1996, \mn@doi [\aaps] {10.1051/aas:1996164}, \href
  {http://adsabs.harvard.edu/abs/1996A%26AS..117..393B} {117, 393}

\bibitem[\protect\citeauthoryear{{Brammer}, {van Dokkum}  \& {Coppi}}{{Brammer}
  et~al.}{2008}]{Brammer2008}
{Brammer} G.~B.,  {van Dokkum} P.~G.,   {Coppi} P.,  2008, \mn@doi [\apj]
  {10.1086/591786}, \href {http://adsabs.harvard.edu/abs/2008ApJ...686.1503B}
  {686, 1503}

\bibitem[\protect\citeauthoryear{{Bruce} et~al.,}{{Bruce}
  et~al.}{2012}]{Bruce2012}
{Bruce} V.~A.,  et~al., 2012, \mn@doi [\mnras]
  {10.1111/j.1365-2966.2012.22087.x}, \href
  {http://adsabs.harvard.edu/abs/2012MNRAS.427.1666B} {427, 1666}

\bibitem[\protect\citeauthoryear{{Bundy}, {Hogg}, {Higgs}, {Nichol}, {Yasuda},
  {Masters}, {Lang}  \& {Wake}}{{Bundy} et~al.}{2012}]{Bundy2012}
{Bundy} K.,  {Hogg} D.~W.,  {Higgs} T.~D.,  {Nichol} R.~C.,  {Yasuda} N.,
  {Masters} K.~L.,  {Lang} D.,   {Wake} D.~A.,  2012, \mn@doi [\aj]
  {10.1088/0004-6256/144/6/188}, \href
  {http://adsabs.harvard.edu/abs/2012AJ....144..188B} {144, 188}

\bibitem[\protect\citeauthoryear{{Bundy} et~al.,}{{Bundy}
  et~al.}{2015}]{Bundy2015}
{Bundy} K.,  et~al., 2015, \mn@doi [\apjs] {10.1088/0067-0049/221/1/15}, \href
  {http://adsabs.harvard.edu/abs/2015ApJS..221...15B} {221, 15}

\bibitem[\protect\citeauthoryear{{Carollo} et~al.,}{{Carollo}
  et~al.}{2013}]{Carollo2013}
{Carollo} C.~M.,  et~al., 2013, \mn@doi [\apj] {10.1088/0004-637X/773/2/112},
  \href {http://adsabs.harvard.edu/abs/2013ApJ...773..112C} {773, 112}

\bibitem[\protect\citeauthoryear{{Cassata} et~al.,}{{Cassata}
  et~al.}{2013}]{Cassata2013}
{Cassata} P.,  et~al., 2013, \mn@doi [\apj] {10.1088/0004-637X/775/2/106},
  \href {http://adsabs.harvard.edu/abs/2013ApJ...775..106C} {775, 106}

\bibitem[\protect\citeauthoryear{{Chabrier}}{{Chabrier}}{2003}]{Chabrier2003}
{Chabrier} G.,  2003, \mn@doi [\pasp] {10.1086/376392}, \href
  {http://adsabs.harvard.edu/abs/2003PASP..115..763C} {115, 763}

\bibitem[\protect\citeauthoryear{{Chilingarian}, {Melchior}  \&
  {Zolotukhin}}{{Chilingarian} et~al.}{2010}]{Chilingarian2010}
{Chilingarian} I.~V.,  {Melchior} A.-L.,   {Zolotukhin} I.~Y.,  2010, \mn@doi
  [\mnras] {10.1111/j.1365-2966.2010.16506.x}, \href
  {http://adsabs.harvard.edu/abs/2010MNRAS.405.1409C} {405, 1409}

\bibitem[\protect\citeauthoryear{{Cimatti} et~al.,}{{Cimatti}
  et~al.}{2004}]{Cimatti2004}
{Cimatti} A.,  et~al., 2004, \mn@doi [\nat] {10.1038/nature02668}, \href
  {http://adsabs.harvard.edu/abs/2004Natur.430..184C} {430, 184}

\bibitem[\protect\citeauthoryear{{Cimatti} et~al.,}{{Cimatti}
  et~al.}{2008}]{Cimatti2008}
{Cimatti} A.,  et~al., 2008, \mn@doi [\aap] {10.1051/0004-6361:20078739}, \href
  {http://adsabs.harvard.edu/abs/2008A%26A...482...21C} {482, 21}

\bibitem[\protect\citeauthoryear{{Citro}, {Pozzetti}, {Moresco}  \&
  {Cimatti}}{{Citro} et~al.}{2016}]{Citro2016}
{Citro} A.,  {Pozzetti} L.,  {Moresco} M.,   {Cimatti} A.,  2016, preprint,
  \href {http://adsabs.harvard.edu/abs/2016arXiv160407826C} {} (\mn@eprint
  {arXiv} {1604.07826})

\bibitem[\protect\citeauthoryear{{Coil} et~al.,}{{Coil}
  et~al.}{2011}]{Coil2011}
{Coil} A.~L.,  et~al., 2011, \mn@doi [\apj] {10.1088/0004-637X/741/1/8}, \href
  {http://adsabs.harvard.edu/abs/2011ApJ...741....8C} {741, 8}

\bibitem[\protect\citeauthoryear{{Colless} et~al.,}{{Colless}
  et~al.}{2001}]{2dFGRS}
{Colless} M.,  et~al., 2001, \mn@doi [\mnras]
  {10.1046/j.1365-8711.2001.04902.x}, \href
  {http://adsabs.harvard.edu/abs/2001MNRAS.328.1039C} {328, 1039}

\bibitem[\protect\citeauthoryear{{Collister} \& {Lahav}}{{Collister} \&
  {Lahav}}{2004}]{Collister2004}
{Collister} A.~A.,  {Lahav} O.,  2004, \mn@doi [\pasp] {10.1086/383254}, \href
  {http://adsabs.harvard.edu/abs/2004PASP..116..345C} {116, 345}

\bibitem[\protect\citeauthoryear{{Courteau} et~al.,}{{Courteau}
  et~al.}{2014}]{Courteau2014}
{Courteau} S.,  et~al., 2014, \mn@doi [Reviews of Modern Physics]
  {10.1103/RevModPhys.86.47}, \href
  {http://adsabs.harvard.edu/abs/2014RvMP...86...47C} {86, 47}

\bibitem[\protect\citeauthoryear{{Cowie}, {Songaila}, {Hu}  \& {Cohen}}{{Cowie}
  et~al.}{1996}]{Cowie1996}
{Cowie} L.~L.,  {Songaila} A.,  {Hu} E.~M.,   {Cohen} J.~G.,  1996, \mn@doi
  [\aj] {10.1086/118058}, \href
  {http://adsabs.harvard.edu/abs/1996AJ....112..839C} {112, 839}

\bibitem[\protect\citeauthoryear{{Crain} et~al.,}{{Crain}
  et~al.}{2015}]{Crain2015}
{Crain} R.~A.,  et~al., 2015, \mn@doi [\mnras] {10.1093/mnras/stv725}, \href
  {http://adsabs.harvard.edu/abs/2015MNRAS.450.1937C} {450, 1937}

\bibitem[\protect\citeauthoryear{{Croom}, {Smith}, {Boyle}, {Shanks},
  {Loaring}, {Miller}  \& {Lewis}}{{Croom} et~al.}{2001}]{2QZ}
{Croom} S.~M.,  {Smith} R.~J.,  {Boyle} B.~J.,  {Shanks} T.,  {Loaring} N.~S.,
  {Miller} L.,   {Lewis} I.~J.,  2001, \mn@doi [\mnras]
  {10.1046/j.1365-8711.2001.04474.x}, \href
  {http://adsabs.harvard.edu/abs/2001MNRAS.322L..29C} {322, L29}

\bibitem[\protect\citeauthoryear{{Daddi} et~al.,}{{Daddi}
  et~al.}{2005}]{Daddi2005}
{Daddi} E.,  et~al., 2005, \mn@doi [\apj] {10.1086/430104}, \href
  {http://adsabs.harvard.edu/abs/2005ApJ...626..680D} {626, 680}

\bibitem[\protect\citeauthoryear{{Damjanov} et~al.,}{{Damjanov}
  et~al.}{2009}]{Damjanov2009}
{Damjanov} I.,  et~al., 2009, \mn@doi [\apj] {10.1088/0004-637X/695/1/101},
  \href {http://adsabs.harvard.edu/abs/2009ApJ...695..101D} {695, 101}

\bibitem[\protect\citeauthoryear{{Damjanov} et~al.,}{{Damjanov}
  et~al.}{2011}]{Damjanov2011}
{Damjanov} I.,  et~al., 2011, \mn@doi [\apjl] {10.1088/2041-8205/739/2/L44},
  \href {http://adsabs.harvard.edu/abs/2011ApJ...739L..44D} {739, L44}

\bibitem[\protect\citeauthoryear{{Damjanov}, {Chilingarian}, {Hwang}  \&
  {Geller}}{{Damjanov} et~al.}{2013}]{Damjanov2013}
{Damjanov} I.,  {Chilingarian} I.,  {Hwang} H.~S.,   {Geller} M.~J.,  2013,
  \mn@doi [\apjl] {10.1088/2041-8205/775/2/L48}, \href
  {http://adsabs.harvard.edu/abs/2013ApJ...775L..48D} {775, L48}

\bibitem[\protect\citeauthoryear{{Damjanov}, {Hwang}, {Geller}  \&
  {Chilingarian}}{{Damjanov} et~al.}{2014}]{Damjanov2014}
{Damjanov} I.,  {Hwang} H.~S.,  {Geller} M.~J.,   {Chilingarian} I.,  2014,
  \mn@doi [\apj] {10.1088/0004-637X/793/1/39}, \href
  {http://adsabs.harvard.edu/abs/2014ApJ...793...39D} {793, 39}

\bibitem[\protect\citeauthoryear{{Damjanov}, {Geller}, {Zahid}  \&
  {Hwang}}{{Damjanov} et~al.}{2015}]{Damjanov2015}
{Damjanov} I.,  {Geller} M.~J.,  {Zahid} H.~J.,   {Hwang} H.~S.,  2015, \mn@doi
  [\apj] {10.1088/0004-637X/806/2/158}, \href
  {http://adsabs.harvard.edu/abs/2015ApJ...806..158D} {806, 158}

\bibitem[\protect\citeauthoryear{{Dark Energy Survey Collaboration}
  et~al.,}{{Dark Energy Survey Collaboration} et~al.}{2016}]{DES2016}
{Dark Energy Survey Collaboration} et~al., 2016, \mn@doi [\mnras]
  {10.1093/mnras/stw641}, \href
  {http://adsabs.harvard.edu/abs/2016MNRAS.460.1270D} {460, 1270}

\bibitem[\protect\citeauthoryear{{Dawson} et~al.,}{{Dawson}
  et~al.}{2013}]{Dawson2013}
{Dawson} K.~S.,  et~al., 2013, \mn@doi [\aj] {10.1088/0004-6256/145/1/10},
  \href {http://adsabs.harvard.edu/abs/2013AJ....145...10D} {145, 10}

\bibitem[\protect\citeauthoryear{{Dekel} \& {Burkert}}{{Dekel} \&
  {Burkert}}{2014}]{Dekel2014}
{Dekel} A.,  {Burkert} A.,  2014, \mn@doi [\mnras] {10.1093/mnras/stt2331},
  \href {http://adsabs.harvard.edu/abs/2014MNRAS.438.1870D} {438, 1870}

\bibitem[\protect\citeauthoryear{{Desai} et~al.,}{{Desai}
  et~al.}{2012}]{Desai2012}
{Desai} S.,  et~al., 2012, \mn@doi [\apj] {10.1088/0004-637X/757/1/83}, \href
  {http://adsabs.harvard.edu/abs/2012ApJ...757...83D} {757, 83}

\bibitem[\protect\citeauthoryear{{Diemand} \& {Moore}}{{Diemand} \&
  {Moore}}{2011}]{Diemand2011}
{Diemand} J.,  {Moore} B.,  2011, \mn@doi [Advanced Science Letters]
  {10.1166/asl.2011.1211}, \href
  {http://adsabs.harvard.edu/abs/2011ASL.....4..297D} {4, 297}

\bibitem[\protect\citeauthoryear{{Drinkwater} et~al.,}{{Drinkwater}
  et~al.}{2010}]{Drinkwater2010}
{Drinkwater} M.~J.,  et~al., 2010, \mn@doi [\mnras]
  {10.1111/j.1365-2966.2009.15754.x}, \href
  {http://adsabs.harvard.edu/abs/2010MNRAS.401.1429D} {401, 1429}

\bibitem[\protect\citeauthoryear{{Eisenstein} et~al.,}{{Eisenstein}
  et~al.}{2011}]{Eisenstein2011}
{Eisenstein} D.~J.,  et~al., 2011, \mn@doi [\aj] {10.1088/0004-6256/142/3/72},
  \href {http://adsabs.harvard.edu/abs/2011AJ....142...72E} {142, 72}

\bibitem[\protect\citeauthoryear{{Fliri} \& {Trujillo}}{{Fliri} \&
  {Trujillo}}{2016}]{Fliri2016}
{Fliri} J.,  {Trujillo} I.,  2016, \mn@doi [\mnras] {10.1093/mnras/stv2686},
  \href {http://adsabs.harvard.edu/abs/2016MNRAS.456.1359F} {456, 1359}

\bibitem[\protect\citeauthoryear{{F{\"o}rster Schreiber} et~al.,}{{F{\"o}rster
  Schreiber} et~al.}{2014}]{ForsterSchreiber2014}
{F{\"o}rster Schreiber} N.~M.,  et~al., 2014, \mn@doi [\apj]
  {10.1088/0004-637X/787/1/38}, \href
  {http://adsabs.harvard.edu/abs/2014ApJ...787...38F} {787, 38}

\bibitem[\protect\citeauthoryear{{Furlong} et~al.,}{{Furlong}
  et~al.}{2015}]{Furlong2015}
{Furlong} M.,  et~al., 2015, preprint, \href
  {http://adsabs.harvard.edu/abs/2015arXiv151005645F} {} (\mn@eprint {arXiv}
  {1510.05645})

\bibitem[\protect\citeauthoryear{{Garilli} et~al.,}{{Garilli}
  et~al.}{2008}]{Garilli2008}
{Garilli} B.,  et~al., 2008, \mn@doi [\aap] {10.1051/0004-6361:20078878}, \href
  {http://adsabs.harvard.edu/abs/2008A%26A...486..683G} {486, 683}

\bibitem[\protect\citeauthoryear{{Gon{\c c}alves}, {Martin},
  {Men{\'e}ndez-Delmestre}, {Wyder}  \& {Koekemoer}}{{Gon{\c c}alves}
  et~al.}{2012}]{Goncalves2012}
{Gon{\c c}alves} T.~S.,  {Martin} D.~C.,  {Men{\'e}ndez-Delmestre} K.,  {Wyder}
  T.~K.,   {Koekemoer} A.,  2012, \mn@doi [\apj] {10.1088/0004-637X/759/1/67},
  \href {http://adsabs.harvard.edu/abs/2012ApJ...759...67G} {759, 67}

\bibitem[\protect\citeauthoryear{{Graham}, {Dullo}  \& {Savorgnan}}{{Graham}
  et~al.}{2015}]{Graham2015}
{Graham} A.~W.,  {Dullo} B.~T.,   {Savorgnan} G.~A.~D.,  2015, \mn@doi [\apj]
  {10.1088/0004-637X/804/1/32}, \href
  {http://adsabs.harvard.edu/abs/2015ApJ...804...32G} {804, 32}

\bibitem[\protect\citeauthoryear{{Heymans} et~al.,}{{Heymans}
  et~al.}{2012}]{Heymans2012}
{Heymans} C.,  et~al., 2012, \mn@doi [\mnras]
  {10.1111/j.1365-2966.2012.21952.x}, \href
  {http://adsabs.harvard.edu/abs/2012MNRAS.427..146H} {427, 146}

\bibitem[\protect\citeauthoryear{{Heywood} et~al.,}{{Heywood}
  et~al.}{2016}]{Heywood2016}
{Heywood} I.,  et~al., 2016, \mn@doi [\mnras] {10.1093/mnras/stw1250}, \href
  {http://adsabs.harvard.edu/abs/2016MNRAS.460.4433H} {460, 4433}

\bibitem[\protect\citeauthoryear{{Hodge}, {Becker}, {White}, {Richards}  \&
  {Zeimann}}{{Hodge} et~al.}{2011}]{Hodge2011}
{Hodge} J.~A.,  {Becker} R.~H.,  {White} R.~L.,  {Richards} G.~T.,   {Zeimann}
  G.~R.,  2011, \mn@doi [\aj] {10.1088/0004-6256/142/1/3}, \href
  {http://adsabs.harvard.edu/abs/2011AJ....142....3H} {142, 3}

\bibitem[\protect\citeauthoryear{{Huertas-Company} et~al.,}{{Huertas-Company}
  et~al.}{2013}]{HuertasCompany2013}
{Huertas-Company} M.,  et~al., 2013, \mn@doi [\mnras] {10.1093/mnras/sts150},
  \href {http://adsabs.harvard.edu/abs/2013MNRAS.428.1715H} {428, 1715}

\bibitem[\protect\citeauthoryear{{Huertas-Company} et~al.,}{{Huertas-Company}
  et~al.}{2015}]{HuertasCompany2015}
{Huertas-Company} M.,  et~al., 2015, \mn@doi [\apj]
  {10.1088/0004-637X/809/1/95}, \href
  {http://adsabs.harvard.edu/abs/2015ApJ...809...95H} {809, 95}

\bibitem[\protect\citeauthoryear{{Ilbert} et~al.,}{{Ilbert}
  et~al.}{2013}]{Ilbert2013}
{Ilbert} O.,  et~al., 2013, \mn@doi [\aap] {10.1051/0004-6361/201321100}, \href
  {http://adsabs.harvard.edu/abs/2013A%26A...556A..55I} {556, A55}

\bibitem[\protect\citeauthoryear{{Jiang} et~al.,}{{Jiang}
  et~al.}{2014}]{Jiang2014}
{Jiang} L.,  et~al., 2014, \mn@doi [\apjs] {10.1088/0067-0049/213/1/12}, \href
  {http://adsabs.harvard.edu/abs/2014ApJS..213...12J} {213, 12}

\bibitem[\protect\citeauthoryear{{Jones} et~al.,}{{Jones} et~al.}{2009}]{6dF}
{Jones} D.~H.,  et~al., 2009, \mn@doi [\mnras]
  {10.1111/j.1365-2966.2009.15338.x}, \href
  {http://adsabs.harvard.edu/abs/2009MNRAS.399..683J} {399, 683}

\bibitem[\protect\citeauthoryear{{Kelvin} et~al.,}{{Kelvin}
  et~al.}{2012}]{Kelvin2012}
{Kelvin} L.~S.,  et~al., 2012, \mn@doi [\mnras]
  {10.1111/j.1365-2966.2012.20355.x}, \href
  {http://adsabs.harvard.edu/abs/2012MNRAS.421.1007K} {421, 1007}

\bibitem[\protect\citeauthoryear{{LSST Science Collaboration} et~al.,}{{LSST
  Science Collaboration} et~al.}{2009}]{LSST2009}
{LSST Science Collaboration} et~al., 2009, preprint, \href
  {http://adsabs.harvard.edu/abs/2009arXiv0912.0201L} {} (\mn@eprint {arXiv}
  {0912.0201})

\bibitem[\protect\citeauthoryear{{LaMassa} et~al.,}{{LaMassa}
  et~al.}{2016}]{LaMassa2016}
{LaMassa} S.~M.,  et~al., 2016, \mn@doi [\apj] {10.3847/0004-637X/817/2/172},
  \href {http://adsabs.harvard.edu/abs/2016ApJ...817..172L} {817, 172}

\bibitem[\protect\citeauthoryear{{Lawrence} et~al.,}{{Lawrence}
  et~al.}{2007}]{Lawrence2007}
{Lawrence} A.,  et~al., 2007, \mn@doi [\mnras]
  {10.1111/j.1365-2966.2007.12040.x}, \href
  {http://adsabs.harvard.edu/abs/2007MNRAS.379.1599L} {379, 1599}

\bibitem[\protect\citeauthoryear{{Lilly} \& {Carollo}}{{Lilly} \&
  {Carollo}}{2016}]{Lilly2016}
{Lilly} S.~J.,  {Carollo} C.~M.,  2016, preprint, \href
  {http://adsabs.harvard.edu/abs/2016arXiv160406459L} {} (\mn@eprint {arXiv}
  {1604.06459})

\bibitem[\protect\citeauthoryear{{Longhetti} et~al.,}{{Longhetti}
  et~al.}{2007}]{Longhetti2007}
{Longhetti} M.,  et~al., 2007, \mn@doi [\mnras]
  {10.1111/j.1365-2966.2006.11171.x}, \href
  {http://adsabs.harvard.edu/abs/2007MNRAS.374..614L} {374, 614}

\bibitem[\protect\citeauthoryear{{Madau} \& {Dickinson}}{{Madau} \&
  {Dickinson}}{2014}]{Madau2014}
{Madau} P.,  {Dickinson} M.,  2014, \mn@doi [\araa]
  {10.1146/annurev-astro-081811-125615}, \href
  {http://adsabs.harvard.edu/abs/2014ARA%26A..52..415M} {52, 415}

\bibitem[\protect\citeauthoryear{{Madau}, {Pozzetti}  \& {Dickinson}}{{Madau}
  et~al.}{1998}]{Madau1998}
{Madau} P.,  {Pozzetti} L.,   {Dickinson} M.,  1998, \mn@doi [\apj]
  {10.1086/305523}, \href {http://adsabs.harvard.edu/abs/1998ApJ...498..106M}
  {498, 106}

\bibitem[\protect\citeauthoryear{{Martin} et~al.,}{{Martin}
  et~al.}{2007}]{Martin2007}
{Martin} D.~C.,  et~al., 2007, \mn@doi [\apjs] {10.1086/516639}, \href
  {http://adsabs.harvard.edu/abs/2007ApJS..173..342M} {173, 342}

\bibitem[\protect\citeauthoryear{{Miller} et~al.,}{{Miller}
  et~al.}{2013}]{Miller2013}
{Miller} L.,  et~al., 2013, \mn@doi [\mnras] {10.1093/mnras/sts454}, \href
  {http://adsabs.harvard.edu/abs/2013MNRAS.429.2858M} {429, 2858}

\bibitem[\protect\citeauthoryear{Miyazaki et~al.,}{Miyazaki
  et~al.}{2012}]{Miyazaki2012}
Miyazaki S.,  et~al., 2012, Hyper Suprime-Cam, \mn@doi{10.1117/12.926844}, \url
  {http://dx.doi.org/10.1117/12.926844}

\bibitem[\protect\citeauthoryear{{Mo}, {Mao}  \& {White}}{{Mo}
  et~al.}{1998}]{Mo1998}
{Mo} H.~J.,  {Mao} S.,   {White} S.~D.~M.,  1998, \mn@doi [\mnras]
  {10.1046/j.1365-8711.1998.01227.x}, \href
  {http://adsabs.harvard.edu/abs/1998MNRAS.295..319M} {295, 319}

\bibitem[\protect\citeauthoryear{{Mooley} et~al.,}{{Mooley}
  et~al.}{2016}]{Mooley2016}
{Mooley} K.~P.,  et~al., 2016, \mn@doi [\apj] {10.3847/0004-637X/818/2/105},
  \href {http://adsabs.harvard.edu/abs/2016ApJ...818..105M} {818, 105}

\bibitem[\protect\citeauthoryear{{Moresco} et~al.,}{{Moresco}
  et~al.}{2013}]{Moresco2013}
{Moresco} M.,  et~al., 2013, \mn@doi [\aap] {10.1051/0004-6361/201321797},
  \href {http://adsabs.harvard.edu/abs/2013A%26A...558A..61M} {558, A61}

\bibitem[\protect\citeauthoryear{{Moster}, {Somerville}, {Newman}  \&
  {Rix}}{{Moster} et~al.}{2011}]{Moster2011}
{Moster} B.~P.,  {Somerville} R.~S.,  {Newman} J.~A.,   {Rix} H.-W.,  2011,
  \mn@doi [\apj] {10.1088/0004-637X/731/2/113}, \href
  {http://adsabs.harvard.edu/abs/2011ApJ...731..113M} {731, 113}

\bibitem[\protect\citeauthoryear{{Muzzin} et~al.,}{{Muzzin}
  et~al.}{2013}]{Muzzin2013}
{Muzzin} A.,  et~al., 2013, \mn@doi [\apjs] {10.1088/0067-0049/206/1/8}, \href
  {http://adsabs.harvard.edu/abs/2013ApJS..206....8M} {206, 8}

\bibitem[\protect\citeauthoryear{{Newman}, {Ellis}, {Treu}  \&
  {Bundy}}{{Newman} et~al.}{2010}]{Newman2010}
{Newman} A.~B.,  {Ellis} R.~S.,  {Treu} T.,   {Bundy} K.,  2010, \mn@doi
  [\apjl] {10.1088/2041-8205/717/2/L103}, \href
  {http://adsabs.harvard.edu/abs/2010ApJ...717L.103N} {717, L103}

\bibitem[\protect\citeauthoryear{{Newman} et~al.,}{{Newman}
  et~al.}{2012}]{DEEP2}
{Newman} J.~A.,  et~al., 2012, preprint, \href
  {http://adsabs.harvard.edu/abs/2012arXiv1203.3192N} {} (\mn@eprint {arXiv}
  {1203.3192})

\bibitem[\protect\citeauthoryear{{Oser}, {Naab}, {Ostriker}  \&
  {Johansson}}{{Oser} et~al.}{2012}]{Oser2012}
{Oser} L.,  {Naab} T.,  {Ostriker} J.~P.,   {Johansson} P.~H.,  2012, \mn@doi
  [\apj] {10.1088/0004-637X/744/1/63}, \href
  {http://adsabs.harvard.edu/abs/2012ApJ...744...63O} {744, 63}

\bibitem[\protect\citeauthoryear{{Papovich} et~al.,}{{Papovich}
  et~al.}{2016}]{Papovich2016}
{Papovich} C.,  et~al., 2016, \mn@doi [\apjs] {10.3847/0067-0049/224/2/28},
  \href {http://adsabs.harvard.edu/abs/2016ApJS..224...28P} {224, 28}

\bibitem[\protect\citeauthoryear{{Peng} et~al.,}{{Peng}
  et~al.}{2010}]{Peng2010}
{Peng} Y.-j.,  et~al., 2010, \mn@doi [\apj] {10.1088/0004-637X/721/1/193},
  \href {http://adsabs.harvard.edu/abs/2010ApJ...721..193P} {721, 193}

\bibitem[\protect\citeauthoryear{{Peralta de Arriba}, {Quilis}, {Trujillo},
  {Cebri{\'a}n}  \& {Balcells}}{{Peralta de Arriba}
  et~al.}{2016}]{PeraltaDeArriba2016}
{Peralta de Arriba} L.,  {Quilis} V.,  {Trujillo} I.,  {Cebri{\'a}n} M.,
  {Balcells} M.,  2016, preprint, \href
  {http://adsabs.harvard.edu/abs/2016arXiv160506503P} {} (\mn@eprint {arXiv}
  {1605.06503})

\bibitem[\protect\citeauthoryear{{Poggianti} et~al.,}{{Poggianti}
  et~al.}{2013}]{Poggianti2013a}
{Poggianti} B.~M.,  et~al., 2013, \mn@doi [\apj] {10.1088/0004-637X/762/2/77},
  \href {http://adsabs.harvard.edu/abs/2013ApJ...762...77P} {762, 77}

\bibitem[\protect\citeauthoryear{{Pozzetti} et~al.,}{{Pozzetti}
  et~al.}{2010}]{Pozzetti2010}
{Pozzetti} L.,  et~al., 2010, \mn@doi [\aap] {10.1051/0004-6361/200913020},
  \href {http://adsabs.harvard.edu/abs/2010A%26A...523A..13P} {523, A13}

\bibitem[\protect\citeauthoryear{{Quilis} \& {Trujillo}}{{Quilis} \&
  {Trujillo}}{2013}]{Quilis2013}
{Quilis} V.,  {Trujillo} I.,  2013, \mn@doi [\apjl]
  {10.1088/2041-8205/773/1/L8}, \href
  {http://adsabs.harvard.edu/abs/2013ApJ...773L...8Q} {773, L8}

\bibitem[\protect\citeauthoryear{{Reis} et~al.,}{{Reis}
  et~al.}{2012}]{Reis2012}
{Reis} R.~R.~R.,  et~al., 2012, \mn@doi [\apj] {10.1088/0004-637X/747/1/59},
  \href {http://adsabs.harvard.edu/abs/2012ApJ...747...59R} {747, 59}

\bibitem[\protect\citeauthoryear{{Rosen} et~al.,}{{Rosen}
  et~al.}{2016}]{Rosen2016}
{Rosen} S.~R.,  et~al., 2016, \mn@doi [\aap] {10.1051/0004-6361/201526416},
  \href {http://adsabs.harvard.edu/abs/2016A%26A...590A...1R} {590, A1}

\bibitem[\protect\citeauthoryear{{Ryan} Jr. et~al.,}{{Ryan}
  et~al.}{2012}]{Ryan2012}
{Ryan} Jr. R.~E.,  et~al., 2012, \mn@doi [\apj] {10.1088/0004-637X/749/1/53},
  \href {http://adsabs.harvard.edu/abs/2012ApJ...749...53R} {749, 53}

\bibitem[\protect\citeauthoryear{{Rykoff} et~al.,}{{Rykoff}
  et~al.}{2014}]{Rykoff2014}
{Rykoff} E.~S.,  et~al., 2014, \mn@doi [\apj] {10.1088/0004-637X/785/2/104},
  \href {http://adsabs.harvard.edu/abs/2014ApJ...785..104R} {785, 104}

\bibitem[\protect\citeauthoryear{{SDSS Collaboration} et~al.,}{{SDSS
  Collaboration} et~al.}{2016}]{SDSSDR13}
{SDSS Collaboration} et~al., 2016, preprint, \href
  {http://adsabs.harvard.edu/abs/2016arXiv160802013S} {} (\mn@eprint {arXiv}
  {1608.02013})

\bibitem[\protect\citeauthoryear{{Schaye} et~al.,}{{Schaye}
  et~al.}{2015}]{Schaye2015}
{Schaye} J.,  et~al., 2015, \mn@doi [\mnras] {10.1093/mnras/stu2058}, \href
  {http://adsabs.harvard.edu/abs/2015MNRAS.446..521S} {446, 521}

\bibitem[\protect\citeauthoryear{{Schmidt}}{{Schmidt}}{1968}]{Schmidt1968}
{Schmidt} M.,  1968, \mn@doi [\apj] {10.1086/149446}, \href
  {http://adsabs.harvard.edu/abs/1968ApJ...151..393S} {151, 393}

\bibitem[\protect\citeauthoryear{{Shen}, {Mo}, {White}, {Blanton}, {Kauffmann},
  {Voges}, {Brinkmann}  \& {Csabai}}{{Shen} et~al.}{2003}]{Shen2003}
{Shen} S.,  {Mo} H.~J.,  {White} S.~D.~M.,  {Blanton} M.~R.,  {Kauffmann} G.,
  {Voges} W.,  {Brinkmann} J.,   {Csabai} I.,  2003, \mn@doi [\mnras]
  {10.1046/j.1365-8711.2003.06740.x}, \href
  {http://adsabs.harvard.edu/abs/2003MNRAS.343..978S} {343, 978}

\bibitem[\protect\citeauthoryear{{Somerville} \& {Dav{\'e}}}{{Somerville} \&
  {Dav{\'e}}}{2015}]{Somerville2015}
{Somerville} R.~S.,  {Dav{\'e}} R.,  2015, \mn@doi [\araa]
  {10.1146/annurev-astro-082812-140951}, \href
  {http://adsabs.harvard.edu/abs/2015ARA%26A..53...51S} {53, 51}

\bibitem[\protect\citeauthoryear{{Strateva} et~al.,}{{Strateva}
  et~al.}{2001}]{Strateva2001}
{Strateva} I.,  et~al., 2001, \mn@doi [\aj] {10.1086/323301}, \href
  {http://adsabs.harvard.edu/abs/2001AJ....122.1861S} {122, 1861}

\bibitem[\protect\citeauthoryear{{Stringer}, {Trujillo}, {Dalla Vecchia}  \&
  {Martinez-Valpuesta}}{{Stringer} et~al.}{2015}]{Stringer2015}
{Stringer} M.,  {Trujillo} I.,  {Dalla Vecchia} C.,   {Martinez-Valpuesta} I.,
  2015, \mn@doi [\mnras] {10.1093/mnras/stv455}, \href
  {http://adsabs.harvard.edu/abs/2015MNRAS.449.2396S} {449, 2396}

\bibitem[\protect\citeauthoryear{{Swetz} et~al.,}{{Swetz}
  et~al.}{2011}]{Swetz2011}
{Swetz} D.~S.,  et~al., 2011, \mn@doi [\apjs] {10.1088/0067-0049/194/2/41},
  \href {http://adsabs.harvard.edu/abs/2011ApJS..194...41S} {194, 41}

\bibitem[\protect\citeauthoryear{{Taylor}, {Franx}, {Glazebrook}, {Brinchmann},
  {van der Wel}  \& {van Dokkum}}{{Taylor} et~al.}{2010}]{Taylor2010}
{Taylor} E.~N.,  {Franx} M.,  {Glazebrook} K.,  {Brinchmann} J.,  {van der Wel}
  A.,   {van Dokkum} P.~G.,  2010, \mn@doi [\apj]
  {10.1088/0004-637X/720/1/723}, \href
  {http://adsabs.harvard.edu/abs/2010ApJ...720..723T} {720, 723}

\bibitem[\protect\citeauthoryear{{Timlin} et~al.,}{{Timlin}
  et~al.}{2016}]{Timlin2016}
{Timlin} J.~D.,  et~al., 2016, \mn@doi [\apjs] {10.3847/0067-0049/225/1/1},
  \href {http://adsabs.harvard.edu/abs/2016ApJS..225....1T} {225, 1}

\bibitem[\protect\citeauthoryear{{Toft} et~al.,}{{Toft}
  et~al.}{2014}]{Toft2014}
{Toft} S.,  et~al., 2014, \mn@doi [\apj] {10.1088/0004-637X/782/2/68}, \href
  {http://adsabs.harvard.edu/abs/2014ApJ...782...68T} {782, 68}

\bibitem[\protect\citeauthoryear{{Tortora} et~al.,}{{Tortora}
  et~al.}{2016}]{Tortora2016}
{Tortora} C.,  et~al., 2016, \mn@doi [\mnras] {10.1093/mnras/stw184}, \href
  {http://adsabs.harvard.edu/abs/2016MNRAS.457.2845T} {457, 2845}

\bibitem[\protect\citeauthoryear{{Trujillo} et~al.,}{{Trujillo}
  et~al.}{2006}]{Trujillo2006}
{Trujillo} I.,  et~al., 2006, \mn@doi [\mnras]
  {10.1111/j.1745-3933.2006.00238.x}, \href
  {http://adsabs.harvard.edu/abs/2006MNRAS.373L..36T} {373, L36}

\bibitem[\protect\citeauthoryear{{Trujillo}, {Cenarro}, {de
  Lorenzo-C{\'a}ceres}, {Vazdekis}, {de la Rosa}  \& {Cava}}{{Trujillo}
  et~al.}{2009}]{Trujillo2009}
{Trujillo} I.,  {Cenarro} A.~J.,  {de Lorenzo-C{\'a}ceres} A.,  {Vazdekis} A.,
  {de la Rosa} I.~G.,   {Cava} A.,  2009, \mn@doi [\apjl]
  {10.1088/0004-637X/692/2/L118}, \href
  {http://adsabs.harvard.edu/abs/2009ApJ...692L.118T} {692, L118}

\bibitem[\protect\citeauthoryear{{Trujillo}, {Ferr{\'e}-Mateu}, {Balcells},
  {Vazdekis}  \& {S{\'a}nchez-Bl{\'a}zquez}}{{Trujillo}
  et~al.}{2014}]{Trujillo2014}
{Trujillo} I.,  {Ferr{\'e}-Mateu} A.,  {Balcells} M.,  {Vazdekis} A.,
  {S{\'a}nchez-Bl{\'a}zquez} P.,  2014, \mn@doi [\apjl]
  {10.1088/2041-8205/780/2/L20}, \href
  {http://adsabs.harvard.edu/abs/2014ApJ...780L..20T} {780, L20}

\bibitem[\protect\citeauthoryear{{Valentinuzzi} et~al.,}{{Valentinuzzi}
  et~al.}{2010a}]{Valentinuzzi2010a}
{Valentinuzzi} T.,  et~al., 2010a, \mn@doi [\apj]
  {10.1088/0004-637X/712/1/226}, \href
  {http://adsabs.harvard.edu/abs/2010ApJ...712..226V} {712, 226}

\bibitem[\protect\citeauthoryear{{Valentinuzzi} et~al.,}{{Valentinuzzi}
  et~al.}{2010b}]{Valentinuzzi2010b}
{Valentinuzzi} T.,  et~al., 2010b, \mn@doi [\apjl]
  {10.1088/2041-8205/721/1/L19}, \href
  {http://adsabs.harvard.edu/abs/2010ApJ...721L..19V} {721, L19}

\bibitem[\protect\citeauthoryear{{Viero} et~al.,}{{Viero}
  et~al.}{2014}]{Viero2014}
{Viero} M.~P.,  et~al., 2014, \mn@doi [\apjs] {10.1088/0067-0049/210/2/22},
  \href {http://adsabs.harvard.edu/abs/2014ApJS..210...22V} {210, 22}

\bibitem[\protect\citeauthoryear{{Vogelsberger} et~al.,}{{Vogelsberger}
  et~al.}{2014}]{Vogelsberger2014}
{Vogelsberger} M.,  et~al., 2014, \mn@doi [\nat] {10.1038/nature13316}, \href
  {http://adsabs.harvard.edu/abs/2014Natur.509..177V} {509, 177}

\bibitem[\protect\citeauthoryear{{Wellons} et~al.,}{{Wellons}
  et~al.}{2015}]{Wellons2015}
{Wellons} S.,  et~al., 2015, \mn@doi [\mnras] {10.1093/mnras/stv303}, \href
  {http://adsabs.harvard.edu/abs/2015MNRAS.449..361W} {449, 361}

\bibitem[\protect\citeauthoryear{{Wellons} et~al.,}{{Wellons}
  et~al.}{2016}]{Wellons2016}
{Wellons} S.,  et~al., 2016, \mn@doi [\mnras] {10.1093/mnras/stv2738}, \href
  {http://adsabs.harvard.edu/abs/2016MNRAS.456.1030W} {456, 1030}

\bibitem[\protect\citeauthoryear{{Whitaker} et~al.,}{{Whitaker}
  et~al.}{2011}]{Whitaker2011}
{Whitaker} K.~E.,  et~al., 2011, \mn@doi [\apj] {10.1088/0004-637X/735/2/86},
  \href {http://adsabs.harvard.edu/abs/2011ApJ...735...86W} {735, 86}

\bibitem[\protect\citeauthoryear{{Whitaker}, {Kriek}, {van Dokkum}, {Bezanson},
  {Brammer}, {Franx}  \& {Labb{\'e}}}{{Whitaker} et~al.}{2012}]{Whitaker2012}
{Whitaker} K.~E.,  {Kriek} M.,  {van Dokkum} P.~G.,  {Bezanson} R.,  {Brammer}
  G.,  {Franx} M.,   {Labb{\'e}} I.,  2012, \mn@doi [\apj]
  {10.1088/0004-637X/745/2/179}, \href
  {http://adsabs.harvard.edu/abs/2012ApJ...745..179W} {745, 179}

\bibitem[\protect\citeauthoryear{{Williams}, {Quadri}, {Franx}, {van Dokkum}
  \& {Labb{\'e}}}{{Williams} et~al.}{2009}]{Williams2009}
{Williams} R.~J.,  {Quadri} R.~F.,  {Franx} M.,  {van Dokkum} P.,   {Labb{\'e}}
  I.,  2009, \mn@doi [\apj] {10.1088/0004-637X/691/2/1879}, \href
  {http://adsabs.harvard.edu/abs/2009ApJ...691.1879W} {691, 1879}

\bibitem[\protect\citeauthoryear{{Williams} et~al.,}{{Williams}
  et~al.}{2015}]{Williams2015}
{Williams} C.~C.,  et~al., 2015, \mn@doi [\apj] {10.1088/0004-637X/800/1/21},
  \href {http://adsabs.harvard.edu/abs/2015ApJ...800...21W} {800, 21}

\bibitem[\protect\citeauthoryear{{Woo}, {Dekel}, {Faber}  \& {Koo}}{{Woo}
  et~al.}{2015}]{Woo2015}
{Woo} J.,  {Dekel} A.,  {Faber} S.~M.,   {Koo} D.~C.,  2015, \mn@doi [\mnras]
  {10.1093/mnras/stu2755}, \href
  {http://adsabs.harvard.edu/abs/2015MNRAS.448..237W} {448, 237}

\bibitem[\protect\citeauthoryear{{Wright} et~al.,}{{Wright}
  et~al.}{2010}]{Wright2010}
{Wright} E.~L.,  et~al., 2010, \mn@doi [\aj] {10.1088/0004-6256/140/6/1868},
  \href {http://adsabs.harvard.edu/abs/2010AJ....140.1868W} {140, 1868}

\bibitem[\protect\citeauthoryear{{Wuyts} et~al.,}{{Wuyts}
  et~al.}{2007}]{Wuyts2007}
{Wuyts} S.,  et~al., 2007, \mn@doi [\apj] {10.1086/509708}, \href
  {http://adsabs.harvard.edu/abs/2007ApJ...655...51W} {655, 51}

\bibitem[\protect\citeauthoryear{{Yoon}, {Weinberg}  \& {Katz}}{{Yoon}
  et~al.}{2011}]{Yoon2011}
{Yoon} I.,  {Weinberg} M.~D.,   {Katz} N.,  2011, \mn@doi [\mnras]
  {10.1111/j.1365-2966.2011.18501.x}, \href
  {http://adsabs.harvard.edu/abs/2011MNRAS.414.1625Y} {414, 1625}

\bibitem[\protect\citeauthoryear{{York} et~al.,}{{York}
  et~al.}{2000}]{York2000}
{York} D.~G.,  et~al., 2000, \mn@doi [\aj] {10.1086/301513}, \href
  {http://adsabs.harvard.edu/abs/2000AJ....120.1579Y} {120, 1579}

\bibitem[\protect\citeauthoryear{{Zolotov} et~al.,}{{Zolotov}
  et~al.}{2015}]{Zolotov2015}
{Zolotov} A.,  et~al., 2015, \mn@doi [\mnras] {10.1093/mnras/stv740}, \href
  {http://adsabs.harvard.edu/abs/2015MNRAS.450.2327Z} {450, 2327}

\bibitem[\protect\citeauthoryear{{Zu} \& {Mandelbaum}}{{Zu} \&
  {Mandelbaum}}{2016}]{Zu2016}
{Zu} Y.,  {Mandelbaum} R.,  2016, \mn@doi [\mnras] {10.1093/mnras/stw221},
  \href {http://adsabs.harvard.edu/abs/2016MNRAS.457.4360Z} {457, 4360}

\bibitem[\protect\citeauthoryear{{de la Rosa}, {La Barbera}, {Ferreras},
  {S{\'a}nchez Almeida}, {Dalla Vecchia}, {Mart{\'{\i}}nez-Valpuesta}  \&
  {Stringer}}{{de la Rosa} et~al.}{2016}]{DelaRosa2016}
{de la Rosa} I.~G.,  {La Barbera} F.,  {Ferreras} I.,  {S{\'a}nchez Almeida}
  J.,  {Dalla Vecchia} C.,  {Mart{\'{\i}}nez-Valpuesta} I.,   {Stringer} M.,
  2016, \mn@doi [\mnras] {10.1093/mnras/stw130}, \href
  {http://adsabs.harvard.edu/abs/2016MNRAS.457.1916D} {457, 1916}

\bibitem[\protect\citeauthoryear{{de la Torre} et~al.,}{{de la Torre}
  et~al.}{2013}]{VIPERS}
{de la Torre} S.,  et~al., 2013, \mn@doi [\aap] {10.1051/0004-6361/201321463},
  \href {http://adsabs.harvard.edu/abs/2013A%26A...557A..54D} {557, A54}

\bibitem[\protect\citeauthoryear{{van Dokkum} et~al.,}{{van Dokkum}
  et~al.}{2008}]{VanDokkum2008}
{van Dokkum} P.~G.,  et~al., 2008, \mn@doi [\apjl] {10.1086/587874}, \href
  {http://adsabs.harvard.edu/abs/2008ApJ...677L...5V} {677, L5}

\bibitem[\protect\citeauthoryear{{van Dokkum} et~al.,}{{van Dokkum}
  et~al.}{2010}]{VanDokkum2010}
{van Dokkum} P.~G.,  et~al., 2010, \mn@doi [\apj]
  {10.1088/0004-637X/709/2/1018}, \href
  {http://adsabs.harvard.edu/abs/2010ApJ...709.1018V} {709, 1018}

\bibitem[\protect\citeauthoryear{{van Dokkum} et~al.,}{{van Dokkum}
  et~al.}{2015}]{VanDokkum2015}
{van Dokkum} P.~G.,  et~al., 2015, \mn@doi [\apj] {10.1088/0004-637X/813/1/23},
  \href {http://adsabs.harvard.edu/abs/2015ApJ...813...23V} {813, 23}

\bibitem[\protect\citeauthoryear{{van der Wel} et~al.,}{{van der Wel}
  et~al.}{2014}]{VanderWel2014}
{van der Wel} A.,  et~al., 2014, \mn@doi [\apj] {10.1088/0004-637X/788/1/28},
  \href {http://adsabs.harvard.edu/abs/2014ApJ...788...28V} {788, 28}

\makeatother
\end{thebibliography}


\bsp	
\label{lastpage}
\end{document}